\documentclass[pra,onecolumn,amsmath,amssymb,superscriptaddress]{revtex4-2}
\usepackage{bm,psfrag,graphicx,setspace}
\usepackage{braket}
\usepackage[left]{lineno}
\usepackage{xcolor}
\selectfont

\begin{document}

\title{Inelastic scattering of transversely structured free electrons from nanophotonic targets: Theory and computation}

\author{Austin G. Nixon}
\affiliation{Department of Chemistry, University of Washington, Seattle, WA 98195} 

\author{Matthieu Chalifour}
\affiliation{Department of Physics, University of Washington, Seattle, WA 98195}

\author{Marc R. Bourgeois}
\affiliation{Department of Chemistry, University of Washington, Seattle, WA 98195}

\author{Michael Sanchez}
\affiliation{Department of Physics, University of Washington, Seattle, WA 98195}

\author{David J. Masiello}
\email{masiello@uw.edu}
\affiliation{Department of Chemistry, University of Washington, Seattle, WA 98195}

\begin{abstract}
Recent advancements in abilities to create and manipulate the electron's transverse wave function within the transmission electron microscope (TEM) and scanning TEM (STEM) have enabled vectorially-resolved electron energy loss (EEL) and gain (EEG) measurements of nanoscale and quantum material responses using pre- and post-selected free electron states. This newfound capability is prompting renewed theoretical interest in quantum mechanical treatments of inelastic electron scattering observables and the information they contain. Here, we present a quantum mechanical treatment of the inelastic scattering of free electrons between pre- and post-selected transverse states with fully-retarded electron--sample interactions for both spontaneous EEL and continuous-wave laser-stimulated EEG measurements. General expressions for the state-resolved energy loss and gain rates are recast in forms amenable to numerical calculation using the method of coupled dipoles. We numerically implement our theory within the $e$-DDA code, and use it to investigate specific examples that highlight its versatility regarding the number, size, geometry, and material composition of the target specimen, as well as its ability to describe matter-wave diffraction from finite nanoscopic targets. 
\end{abstract}

\maketitle

\section{Introduction}
Leveraging recent instrumental advancements in energy monochromation and aberration correction, inelastic scattering of free electrons has become an effective technique to spectroscopically characterize and image atomic and molecular \cite{Lassettre1959-vl, Lassettre2004-gq, PhysRevLett.105.053202}, biological \cite{Adrian1984-wp, Rez2016-ko, Fernandez-Leiro2016-hi}, solid state \cite{Hachtel2019-fz, PhysRevLett.131.116602, PhysRevLett.117.256101, Lagos2017-rn, Senga2019-pe}, and nanophotonic \cite{Polman2019, Auad2023} systems with unprecedented spatial resolution. Simultaneously, optical spectroscopies and microscopies based upon the absorption, scattering, extinction, and emission of electromagnetic waves continue to be indispensable tools used to probe the same systems, albeit with spatial resolution limitations imposed by the optical diffraction limit. Light-based spectromicroscopies can often be enhanced by taking advantage of optical selection rules stemming from the intrinsic linear and spin angular momentum degrees of freedom of the photon \cite{PhysRevA.71.055401}. In addition to the polarization degrees of freedom arising from the intrinsic spin angular momentum, photons can also be prepared in specific orbital angular momentum (OAM) states defined by the azimuthal phase $e^{i \ell \phi}$ \cite{PhysRevA.45.8185}. Due to the helical nature of their spiralling phase fronts, light beams characterized by such a quantized topological charge $\ell$ are commonly referred to as optical vortex or twisted light beams \cite{Molina-Terriza2007-zp, Harris2015-an}. Motivated in part by the infinite dimensional Hilbert space offered by the OAM basis \cite{Yao2011-eq}, the ability to prepare \cite{PhysRevLett.131.183801}, sort \cite{berkhout2010efficient}, and measure \cite{PhysRevLett.104.020505} these optical OAM states has driven applications in quantum information science using photons with quantized azimuthal and radial labels \cite{PhysRevA.89.013829, Harris2015-an, PhysRevLett.119.263602, Zia2023-ao, PhysRevApplied.20.054027, PhysRevA.108.052612}. Moreover, transverse sculpting of the radiation field in general, has lead to the development of new gauge transformations, such as the twisted light gauge \cite{PhysRevA.91.033808, PhysRevA.95.012106}, construction of free space optical skyrmionic beams \cite{PhysRevA.102.053513}, excitations of forbidden transitions in atomic isotopes \cite{PhysRevA.108.043513}, and the use of vortex $\gamma$ ray photons to selectively probe high energy resonances in photonuclear reactions \cite{PhysRevLett.131.202502}.  

Unlike photons, electrons prepared and measured in currently available TEMs, STEMs, or ultrafast TEMs (UTEMs) are accurately described by the spinless free particle Schr\"{o}dinger equation and consequently lack intrinsic polarization degrees of freedom \cite{Harris2015-an}. Despite this limitation, linear-momentum-based selection rules based on quantum mechanical treatments of the inelastic scattering process have been long understood and exploited in core-loss EEL spectroscopy \cite{Kohl1985-fv, Muller1995-zb, Yuan1997-hj, Schattschneider2005-bo} and have enabled measurements of magnetic circular dichroism \cite{Hitchcock1993-vs, Hebert2003-mn, Schattschneider2006-fy, PhysRevB.85.134422, Muto2014-mg, PhysRevLett.113.145501}, characterization of site-specific defects in atomic crystals \cite{PhysRevLett.131.186202}, as well as visualization of the electromagnetic fields of atomic-scale systems \cite{muller2014atomic}. Inspired by the creation and manipulation of optical vortex states, developments of techniques for shaping the transverse phase profile \cite{PhysRevX.12.031043} and OAM content of free electrons via holographic masks \cite{verbeeck2010production, mcmorran2011electron, PhysRevLett.114.034801}, spiral phase plates \cite{uchida2010generation}, and shaped laser pulses \cite{vanacore2019ultrafast, Madan2022-dr} has been at the forefront of low-loss EEL spectroscopy ($\lesssim 50$ eV) \cite{de2010optical}. Furthermore, borrowing ideas from quantum optics, the preparation of free electron qubits carrying information in the form of quantized energy or OAM states using laser pulses \cite{wang2020coherent, PhysRevResearch.3.043033, PhysRevX.13.031001}, holographic masks, or spiral phase plates \cite{Schachinger2021-co, Loffler2022-lh, L_ffler_2023}, has driven the continued development of free electrons as holders and propagators of quantum information. In parallel, the ability to generate phase-structured incident electron states and sort them based upon their OAM content \cite{tavabi2021experimental} has fueled numerous investigations of inelastic electron scattering between states with pre- and post-selected transverse phase profiles in the low-loss regime \cite{asenjo2014dichroism, ugarte2016controlling, guzzinati2017probing, cai2018efficient, zanfrognini2019orbital, PhysRevLett.122.053901, lourencco2021optical, bourgeois2022polarization,bourgeois2023optical, Aguilar2023-zp}. 

In this paper, we expound upon a recently introduced theoretical framework describing the fully retarded inelastic scattering of phase-shaped free electron beams in the electron microscope \cite{bourgeois2023optical}. Specifically, with emphasis placed on the low-loss regime, we investigate the theory of transversely phase-shaped EEL spectroscopy in both narrow beam and wide field limits and laser-stimulated EEG spectroscopy in the narrow beam limit. Sec. \ref{theory_IES} presents a derivation of the energy-resolved inelastic electron scattering rate, including both EEL and EEG processes. Transversely structured free electron states that can be prepared within currently available TEMs and STEMs are subsequently reviewed in Sec. \ref{electron_states_section}. Section \ref{sec_Jfi} introduces the transition current density associated with transitions between such states. Expressions for EEL and EEG observables are derived in Sec. \ref{sec_observables} for both wide field and focused electron beams, including those with nonuniform transverse phase-structure, such as twisted electron beams. Numerical implementation of the presented inelastic electron scattering theory based on the method of coupled dipoles is described in Sec. \ref{sec_numerical}, including EEL and EEG probabilities and the EEL double differential scattering cross section (DDCS). Numerical calculations are presented for prototypical nanophotonic systems, underscoring particular advantages of our numerical approach, including its: (1) flexibility regarding target size, shape, composition, and number, (2) facile extension to accommodate arbitrary initial and final free electron transverse states, and (3) ability to capture signatures of matter-wave diffraction and interference arising from scattering from individual and multiple nanoscale targets. Gaussian units are used throughout.

\section{Inelastic free electron scattering: State- and energy-resolved EEL and EEG rates}\label{theory_IES}

Here we review the retarded theory of inelastic electron scattering for the calculation of EEL and laser-stimulated EEG processes \cite{de2008probing, de2008electron, asenjo2013plasmon, de2010optical, liu2019continuous, das2019stimulated, bourgeois2022nanometer, bourgeois2022polarization, bourgeois2023optical}. The target material is described through its bound electromagnetic responses characterized by dielectric function $\xi(\omega)$, while the electron probe is examined for both delocalized and localized electron wave functions. The light-matter potential governing such interactions takes the form
\begin{equation}
    \hat{V} = \frac{e}{2mc}\Big( \hat{\mathbf{A}} \cdot \hat{\mathbf{p}} + \hat{\mathbf{p}} \cdot \hat{\mathbf{A}} \Big)
    \label{int_potential}
\end{equation}
under minimal coupling in the generalized Coulomb gauge \cite{glauber1991quantum} $\nabla \cdot \xi(\mathbf{x}) \hat{\mathbf{A}}(\mathbf{x},t) = 0$, where the $\hat{\rho}\hat{\Phi}$ term makes no contribution in the absence of free charges describing the target. The probing electron's charge and mass are $-e$ and $m$, and $c$ is the speed of light in vacuum. $\hat{\mathbf{A}}$ and $\hat{\mathbf{p}}$ are quantum mechanical operators for the vector potential of the target and linear momentum of the free electron probe, respectively.

\begin{figure}
    \centering   
\includegraphics{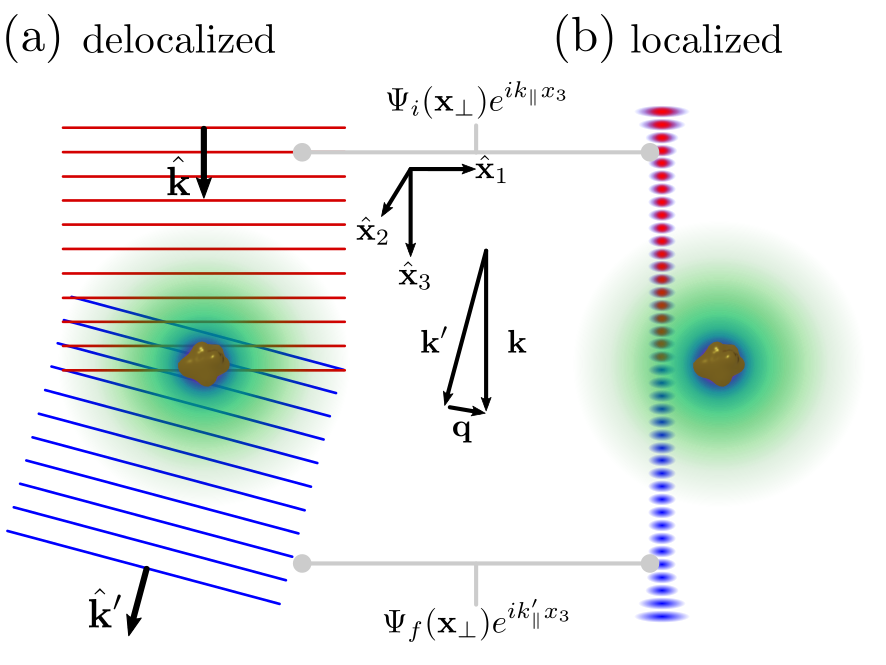}
    \caption{Scheme showing the inelastic scattering of transversely delocalized and localized free electrons with wave functions of the separable form $\Psi({\bf x}_\perp)e^{ik_\parallel x_3}.$ (a) Initial electron plane wave (red) with wave vector $\mathbf{k}=({\bf k}_\perp,k_\parallel)$ interacting with the vector potential (shaded green) of a target specimen. Following the interaction, the electron is post-selected for the plane wave state $\mathbf{k}'=({\bf k}'_\perp,k'_\parallel)$ (blue). The momentum recoil wave vector associated with this transition is $\mathbf{q} = \mathbf{k} - \mathbf{k}'$. (b) A transversely localized electron beam with pre-selected transverse wave function $\Psi_i(\mathbf{x}_{\perp})$ (red) interacting with the same target and post-selected in $\Psi_f(\mathbf{x}_{\perp})$ (blue).}
    \label{F1}
\end{figure}
Owing to the inherently weak nature of electron-photon coupling, the inelastic scattering probability can be obtained using first order time-dependent perturbation theory. The transition rate describing the scattering of an electron from initial state $|\psi_i \rangle$ to final state $| \psi_f \rangle$ while simultaneously exciting/de-exciting the target from initial state $| \Phi_\nu \rangle$ to final state $ |\Phi_{\nu'} \rangle$ is given by
\begin{equation}
    w_{fi} = \frac{2 \pi}{\hbar} \sum_{\nu\nu'}\Big| \langle \Phi_{\nu'}| \langle \psi_f |  \hat{V} | \psi_i \rangle |\Phi_{\nu} \rangle \Big|^{2} \delta(E_f - E_i),
    \label{transition_rate}
\end{equation}
where $E_i=\hbar \varepsilon_i +\hbar\omega_{\nu}$ and $E_f=\hbar \varepsilon_f + \hbar\omega_{\nu'}$ are the initial and final energies of the composite system, $\hbar \varepsilon_{i/f} = \gamma_{i/f}mc^2$ represents the initial/final energy of the probing electron, and $\gamma_{i/f} = \big[1 - (v_{i/f}/ c)^2\big]^{-1/2}$ are the initial/final Lorentz contraction factors. Sec. \ref{electron_states_section} details specific TEM and STEM electron states, including those that are phase-shaped transversely to their propagation direction (see Fig. \ref{F1}), but for now we remain agnostic to their identity. After some algebra, the transition matrix elements of the scattering potential in Eq. (\ref{int_potential}) can be expressed as
\begin{equation}
    \begin{aligned}
        V_{\nu' f \nu i}(t) & =  \langle \Phi_{\nu'}| \langle \psi_f |  \hat{V}(t) | \psi_i \rangle |\Phi_{\nu} \rangle \\
        & = \langle \Phi_{\nu'} | -\frac{1}{c}\int d\mathbf{x} \, \mathbf{A}(\mathbf{x},t) \cdot \mathbf{J}_{fi}(\mathbf{x},t) | \Phi_{\nu} \rangle,
    \end{aligned}
\label{matrix_elements_V}
\end{equation}
where $\mathbf{J}_{fi}(\mathbf{x},t) =\mathbf{J}_{fi}(\mathbf{x})e^{-i(\varepsilon_{i}-\varepsilon_{f}) t}$ is the transition current density \cite{bourgeois2023optical} with 
\begin{equation}
    \mathbf{J}_{fi}(\mathbf{x}) = \frac{i\hbar e}{2m}\big[\psi_f^{*}(\mathbf{x}) \nabla  \psi_i (\mathbf{x}) - \psi_i(\mathbf{x}) \nabla \psi_{f}^{*}(\mathbf{x})\big]
    \label{j_fi}
\end{equation} 
defined in terms of the probe scattering states $\psi_i(\mathbf{x}, t) = \psi_i(\mathbf{x})e^{-i\varepsilon_{i} t}$ and $\psi_f(\mathbf{x}, t) = \psi_f(\mathbf{x})e^{-i\varepsilon_{f} t}$. By continuity, i.e.,  $\nabla \cdot \mathbf{J}_{fi}(\mathbf{x},t)=-\dot{\rho}_{fi}(\mathbf{x},t)$, $\mathbf{J}_{fi}(\mathbf{x}, t)$ is connected to the transition charge density $\rho_{fi}(\mathbf{x},t) = -e\psi_{f}^{*}(\mathbf{x},t)\psi_{i}(\mathbf{x},t)$.

EEL and EEG scattering rates each derive from Eq. (\ref{matrix_elements_V}), but involve different vector potentials of distinct physical origin. In the case of EEL, solving the vector Helmholtz equation on the domain of the target produces a set of eigenmode functions $\mathbf{f}_{\nu}(\mathbf{x})$ and associated eigenmode frequencies $\omega_\nu$ \footnote{
As presented, the eigenmode expansion of the vector potential is rigorously correct only for lossless dielectric cavities. Losses can be incorporated from the outset by instead expanding the vector potential onto the basis of quasinormal modes \cite{PhysRevA.92.053810, Ge2016-xq, PhysRevLett.122.213901}. The resulting EEL rate expressions exhibit equivalent dependence on the Green's dyadic irrespective of the presence or absence of losses}, which serve as a basis to expand the target vector potential \cite{glauber1991quantum}. In terms of this set $\{\mathbf{f}_{\nu}(\mathbf{x})\}_\nu$, $\mathbf{A}(\mathbf{x},t) = \sum_{\nu} [ \mathbf{A}_{\nu}^{(+)}(\mathbf{x}) a_{\nu} e^{-i\omega_\nu t} + \mathbf{A}_{\nu}^{(-)}(\mathbf{x}) a_{\nu}^\dagger e^{i\omega_\nu t}]$, with $\mathbf{A}_{\nu}^{(+)}(\mathbf{x}) = c\sqrt{ 2\pi \hbar /  \omega_\nu}\mathbf{f}_{\nu}(\mathbf{x})$, $\mathbf{A}_{\nu}^{(-)}(\mathbf{x}) = c\sqrt{ 2\pi \hbar /  \omega_\nu}\mathbf{f}_{\nu}^{*}(\mathbf{x})$, and $[\mathbf{A}_{\nu}^{(+)}]^\dagger=\mathbf{A}_{\nu}^{(-)}$, where $a_{\nu}^\dagger$ ($a_{\nu}$) are creation (annihilation) operators responsible for inducing optical excitations (de-excitations) in the $\nu$th target mode. In the case of EEG, the target vector potential can still be expanded onto the $\{\mathbf{f}_{\nu}(\mathbf{x})\}_\nu$ basis, however, it is the response vector potential induced by a stimulating laser field. In either case, Eq. (\ref{matrix_elements_V}) becomes
\begin{widetext}
\begin{equation}
      V_{\nu'f\nu i}(t) = -\frac{1}{c} \Big[ \sqrt{n_{\nu}} \delta_{{\nu'}, \nu-1} \int d\mathbf{x} \,  \mathbf{A}_{\nu}^{(+)}(\mathbf{x}) \cdot \mathbf{J}_{fi}(\mathbf{x})e^{-i\omega_{\nu} t}+ \sqrt{n_{\nu}+1} \delta_{\nu', \nu+1}\int d\mathbf{x} \,  \mathbf{A}_{\nu}^{(-)}(\mathbf{x}) \cdot \mathbf{J}_{fi}(\mathbf{x}) e^{i\omega_{\nu} t}\Big]e^{-i(\varepsilon_{i}-\varepsilon_{f}) t}, 
\label{V_longform}
\end{equation}
\end{widetext}
where $n_{\nu}$ is the occupancy of the $\nu$th target mode. 
When carrying out the derivations of inelastic electron scattering processes, the first and second terms in Eq. (\ref{V_longform}) correspond to EEG and EEL events, respectively. The energy conserving delta function in Eq. (\ref{transition_rate}), when written in the following equivalent forms
\begin{equation}
  \delta(E_f - E_i)= \frac{1}{\hbar}\int d\omega \begin{cases}
    \delta(\omega-\varepsilon_{if}) \delta(\omega - \omega_{\nu'\nu}) \textrm{ \ (EEL)}\\
    \delta(\omega + \varepsilon_{if}) \delta(\omega - \omega_{\nu\nu'}) \textrm{ \ (EEG)},
  \end{cases}
  \label{energy_cons}
\end{equation}
with $\varepsilon_{if} = \varepsilon_{i} - \varepsilon_{f}$ and $\omega_{\nu\nu'} = \omega_{\nu} - \omega_{\nu'}$, will aid in this connection to EEL and EEG.

In EEL events, the probing electron may transfer energy to and retrieve energy from any of the target modes which must be summed over to account for all such loss processes where the electron acts as both pump and probe. After summing over target states $\nu,\nu'$, the EEL scattering rate becomes
\begin{widetext}
\begin{equation}
    w_{fi}^{\textrm{loss}} = \frac{2 \pi}{\hbar c^2}\sum_{\nu}\int d\mathbf{x} d\mathbf{x}' \, \mathbf{J}_{fi}^{*}(\mathbf{x}) \cdot \mathbf{A}_{\nu}^{(+)}(\mathbf{x}) \mathbf{A}_{\nu}^{(-)}(\mathbf{x}') \cdot \mathbf{J}_{fi}(\mathbf{x}') \delta(E_f - E_i),
    \label{lossrate}
\end{equation}
\end{widetext}
while, in the case of EEG, the stimulating laser field is taken to populate the specific target state $\nu$ leaving only a sum over final target states $\nu'$ to be performed, resulting in
\begin{widetext}
\begin{equation}
    w_{fi}^{\textrm{gain}} = \frac{2 \pi}{\hbar c^2} \int d\mathbf{x} d\mathbf{x}' \, \mathbf{J}_{fi}^{*}(\mathbf{x}) \cdot \mathbf{A}_{\nu}^{(-)}(\mathbf{x}) \mathbf{A}_{\nu}^{(+)}(\mathbf{x}') \cdot \mathbf{J}_{fi}(\mathbf{x}') \delta(E_f - E_i).
    \label{gainrate}
\end{equation}
\end{widetext}
Again note that the target vector potentials $\mathbf{A}_{\nu}^{(\pm)}(\mathbf{x})$ in the EEG rate in Eq. (\ref{gainrate}) are understood to originate in response to external laser stimulation and are not induced by the probing electron's transition current $\mathbf{J}_{fi}(\mathbf{x})$.

From Eqs. (\ref{energy_cons}) and (\ref{lossrate}), together with the relationship $w_{fi} = \int d\omega \, w_{fi}(\omega)$, the frequency-resolved EEL rate $w^{\textrm{loss}}_{fi}(\omega)$ can be derived. By expressing the target vector potential in terms of mode functions, i.e., $\mathbf{A}_{\nu}^{(+)}(\mathbf{x}) \mathbf{A}_{\nu}^{(-)}(\mathbf{x}') = (2\pi\hbar c^2/ \omega_{\nu})\mathbf{f}_{\nu}(\mathbf{x}) \mathbf{f}_{\nu}^{*}(\mathbf{x}')$, Eq. (\ref{lossrate}) can be written in terms of the target's electromagnetic Green's tensor, $\tensor{\mathbf{G}}(\mathbf{x}, \mathbf{x}', \omega)=\sum_{\nu} \mathbf{f}_{\nu}(\mathbf{x}) \mathbf{f}_{\nu}^{*}(\mathbf{x}')/(\omega^2 - \omega_{\nu}^2 + i0^{+})$. More specifically, the EEL rate is formulated in terms of the imaginary part of the Green's dyadic, $\textrm{Im}\big\{ \tensor{\mathbf{G}}(\mathbf{x}, \mathbf{x}', \omega) \big\} = -\sum_{\nu} (\pi /  2\omega_{\nu}) \mathbf{f}_{\nu}(\mathbf{x}) \mathbf{f}_{\nu}^{*}(\mathbf{x}')\delta(\omega - \omega_{\nu'\nu})$. As a result, the state- and frequency-resolved EEL transition rate then becomes
\begin{widetext}
\begin{equation}
\begin{split}
    w_{fi}^{\textrm{loss}}(\omega)  &= -\frac{8 \pi}{\hbar} \int d\mathbf{x} d\mathbf{x}' \, \mathbf{J}_{fi}^{*}(\mathbf{x}) \cdot \textrm{Im}\Big[\tensor{\mathbf{G}}(\mathbf{x}, \mathbf{x}', \omega) \Big]\cdot \mathbf{J}_{fi}(\mathbf{x}') \delta(\omega - \varepsilon_{if})\\
    &= -\frac{8 \pi}{\hbar} \int d\mathbf{x} d\mathbf{x}' \, \textrm{Im}\Big[ \mathbf{J}_{fi}^{*}(\mathbf{x}) \cdot \tensor{\mathbf{G}}(\mathbf{x}, \mathbf{x}', \omega) \cdot \mathbf{J}_{fi}(\mathbf{x}') \Big]\delta(\omega - \varepsilon_{if}),
    \label{w_fi_loss}
    \end{split}
\end{equation} 
\end{widetext}
where the bottom line holds for reciprocal media characterized by $\tensor{\mathbf{G}}(\mathbf{x}, \mathbf{x}', \omega) = \tensor{\mathbf{G}}^T(\mathbf{x}', \mathbf{x}, \omega)$. Note that for general phase-shaped EEL processes described by Eq. (\ref{w_fi_loss}), the transition current density can point arbitrarily in 3D space and is not restricted to lie along the TEM axis. 

Alternatively, for the case of laser-stimulated EEG, Eqs. (\ref{energy_cons}) and (\ref{gainrate}) determine the frequency-resolved EEG rate. The positive (negative) frequency portion of the target's laser-induced response field can be expressed in terms of its induced vector potential as $\mathbf{E}_{\nu}^{(\pm)}(\mathbf{x}) =  (\pm i\omega_{\nu}/c) \, \mathbf{A}_{\nu}^{(\pm)}(\mathbf{x})$. When the stimulating laser excites a coherent state of the target $|\alpha_\nu \rangle$, the frequency resolved EEG rate
\begin{equation}
    w_{fi}^{\textrm{gain}}(\omega) = 2\pi \bigg( \frac{|\alpha_\nu|}{\hbar \omega_{\nu}}\bigg)^2 \Big| \int d\mathbf{x} \, \mathbf{E}_{\nu}^{(+)}(\mathbf{x}) \cdot \mathbf{J}_{fi}(\mathbf{x}) \Big|^2 \delta(\omega+\varepsilon_{if})\delta(\omega - \omega_{\nu\nu'})
\label{w_fi_gain_coherent}
\end{equation}
is proportional to the volume integral of the 3D vector transition current density $\mathbf{J}_{fi}(\mathbf{x})$ projected onto the laser-induced electric field $\mathbf{E}_{\nu}^{(+)}(\mathbf{x})$ of the target. For simplicity, the applied monochromatic continuous-wave laser field is chosen to couple to the target's $\nu$th excited state only. In the low photon occupancy limit ($|\alpha_\nu| \approx 1$), this coherent state description yields
\begin{widetext}
\begin{equation}
    w_{fi}^{\textrm{gain}}(\omega) = 2\pi\bigg( \frac{1}{\hbar \omega_\nu} \bigg)^2  \bigg| \int d\mathbf{x} \, \mathbf{E}_{\nu}^{(+)}(\mathbf{x}) \cdot \mathbf{J}_{fi}(\mathbf{x}) \bigg|^2 \delta(\omega+\varepsilon_{if})\delta(\omega - \omega_{\nu\nu'}),
    \label{w_fi_gain}
\end{equation}
\end{widetext}
which has been stated previously \cite{bourgeois2022polarization}.
As will be shown in Section \ref{narrow_beam_section}, if the appropriate choices for the initial and final electron states are made, Eqs. (\ref{w_fi_loss}) and (\ref{w_fi_gain}) reduce to the conventional EEL and EEG probabilities found in the literature \cite{de2008electron, de2010optical, asenjo2013plasmon, bourgeois2022polarization}, but, as expressed here, are generalized to potentially describe polarized EEL and EEG measurements where the wave function of the probing electron is phase-structured in the plane orthogonal to its motion. Therefore, the approach producing Eqs. (\ref{w_fi_loss}), (\ref{w_fi_gain_coherent}), and (\ref{w_fi_gain}) casts phase-shaped EEL and EEG interactions both within the same framework and on equal footing.

\section{Transversely phase-structured free electron states}\label{electron_states_section} 
\begin{figure*}
    \centering   
\includegraphics{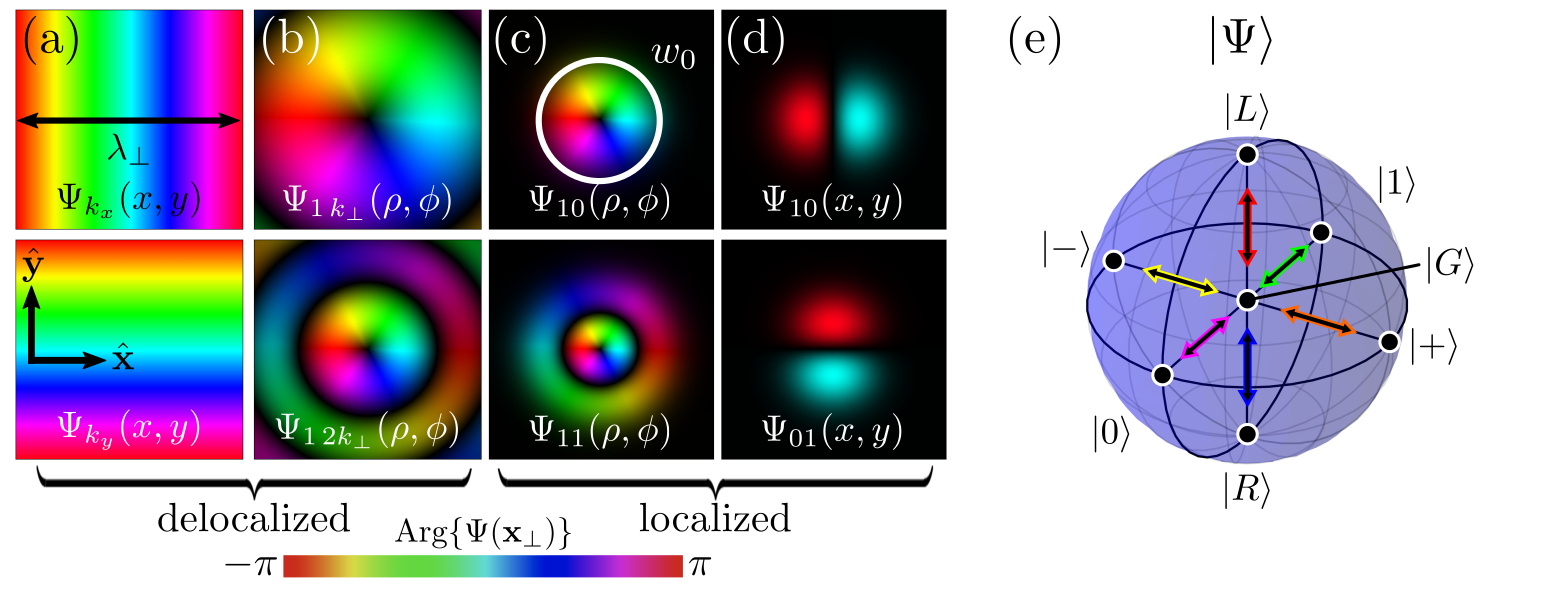}
    \caption{Transverse free electron states. Wave functions $\Psi(x, y)$ are visualized in the $z=0$ plane with position-dependent phase indicated by the color of the respective intensity profiles. Black represents low wave function density. (a) Plane wave states with $\mathbf{k}_{\perp}$ along $\hat{\mathbf{x}}$ (top) and $\hat{\mathbf{y}}$ (bottom). (b) Bessel beam states $\Psi_{\ell k_\perp}(\rho, \phi)$ with $\ell = 1$, and radial wave number $k_{\perp}$ (top) and $2k_{\perp}$ (bottom). (c) LG states $\Psi_{\ell p}(\rho, \phi)$ with $\ell = 1$, and $p = 0$ (top) and $p = 1$ (bottom). (d) HG states $\Psi_{nm}(x, y)$ with $n = 1$ and $m = 0$ (top), and $n = 0$ and $m = 1$ (bottom). (e) In the $\ell = \pm1$ OAM subspace, the first order LG states $|L/R\rangle$ can be used as basis states spanning the Bloch sphere with antipodal points $\big\{ | 0\rangle , |1\rangle\}$, $\big\{ | +\rangle , |-\rangle\},$ and $\big\{ | L\rangle , |R\rangle\}$; states with constant spatial phase are defined as $|G\rangle$ and placed at the sphere center. Colored arrows represent transitions between $|G\rangle$ and the states at the labeled antipodal points.}
    \label{F2}
\end{figure*}

This section introduces various spinless free electron states relevant to the forthcoming discussion of phase-structured EEL and EEG measurements in TEM and STEM instruments. The states are: (i) energy eigenstates and thereby separable into spatial and temporal parts as $\psi(\mathbf{x}, t) = \psi(\mathbf{x})e^{-i\varepsilon t}$ and (ii) separable within an orthogonal coordinate system $\mathbf{x} = (x_1, x_2, x_3)$ into transverse and longitudinal functions $\psi (\mathbf{x}) = \Psi(x_{1}, x_2)e^{ik_\parallel x_3}$. The electron wave functions are translationally invariant along the TEM axis, defined as $x_3 \equiv z$. Transversely delocalized states are investigated first, beginning with plane wave and vortex Bessel beam states originating as separable solutions in the Cartesian and cylindrical coordinate systems, respectively. Subsequently, transversely localized and nondiffracting wave functions, including Hermite-Gauss (HG) and twisted electron Laguerre-Gauss (LG) states, are presented. 

Plane wave solutions are separable in Cartesian coordinates with well-defined linear momentum $\mathbf{p} = \hbar \mathbf{k}$. The spatial wave function describing a free electron in the 3D Cartesian space ${\bf x}=(x,y,z)$ is
\begin{equation}
    \psi_{\mathbf{k}}(x,y,z) = \bigg(\frac{1}{ \sqrt{L}}\bigg)^3 e^{i \mathbf{k} \cdot \mathbf{x}},
    \label{wfxn_epw}
\end{equation}
where $\mathbf{k}=({\bf k}_\perp,k_\parallel)$ is the wave vector and $L$ is the box quantization length. Phase plots of two transverse plane wave states, $\Psi_{k_x}(x,y)$ and $\Psi_{k_y}(x,y)$, with orthogonal wave vectors $\mathbf{k}_{\perp}= k_x\hat{\mathbf{x}}$ and $\mathbf{k}_{\perp} = k_y\hat{\mathbf{y}}$ and corresponding transverse wavelengths $\lambda_{\perp x}=2\pi/k_x$ and $\lambda_{\perp y}=2\pi/k_y$, and are displayed in the upper and lower panels of Fig. \ref{F2}(a) for $\lambda_{\perp x} = \lambda_{\perp y} = \lambda_{\perp}$, respectively. Electrons can also be prepared in coherent superposition states, with one example of such a state being $\psi_{k_xk_y}^{\chi}(x,y,z) = L^{-3/2}\big[e^{ik_x x} + e^{ik_y y}e^{i\chi}\big]e^{ik_{\parallel} z}/\sqrt{2}$, where the relative phase $\chi$ between the two orthogonal electron wave vector components $k_x$ and $k_y$ can take values $0 \leq \chi \leq 2\pi$ \cite{Hitchcock1993-vs, Hebert2003-mn, Schattschneider2006-fy}.

When expressed in the cylindrical coordinate system defined by ${\bf x}=(\rho,\phi, z)$, the separable solutions are nondiffracting Bessel waves of the form \cite{PhysRevX.2.041011, PhysRevA.89.032715, van2015inelastic, PhysRevX.4.011013},
\begin{equation}
    \psi_{\ell k_{\perp}}(\rho, \phi ,z) \propto J_{|\ell|}(k_{\perp} \rho) e^{i \ell \phi }e^{i k_\parallel z}.
    \label{wfxn_Bessel}
\end{equation}
Here, $J_{|\ell|}(k_{\perp} \rho)$ are Bessel functions of the first kind, $\ell$ is the azimuthal quantum number of the cylindrically symmetric state, while $k_{\perp}$ is the radial wave vector component and $k_{\parallel}$ is the longitudinal wave vector component, respectively \cite{bliokh2017theory}. Such states are eigenstates of the $z$-component of the OAM operator $\hat{L}_z = -i\hbar \partial/\partial{\phi }$ with eigenvalue $\ell\hbar$. Specific examples of Bessel states are shown in Fig. \ref{F2}(b) for two different values of $k_{\perp}$. 

Transversely localized free electron states can be constructed within the paraxial approximation to the Schr\"{o}dinger equation, where the electron momentum along the TEM axis is much greater than its transverse momentum, i.e., $k_{\perp} \ll |{\bf k}|$. Since they are eigenstates of the ${\hat L}_z$ operator, LG states carry a quantized azimuthal component $\ell$. In the nondiffracting, i.e., collimated, limit defined by infinite Rayleigh range, the LG wave functions have the form of quantized Landau states \cite{PhysRevX.2.041011, bliokh2017theory} given by
\begin{widetext}
\begin{equation}
    \psi_{\ell p}(\rho, \phi ,z) = \frac{\big(\sqrt{2} \rho/ w_0 \big)^{|\ell|}}{w_0\sqrt{L}} \sqrt{ \frac{2p!}{\pi( |\ell| + p)!} } \textrm{L}_{p}^{|\ell|} \bigg( \frac{2\rho^2}{w^{2}_0} \bigg) \, e^{-\frac{\rho^2}{w_0^{2}}}e^{i\ell \phi  }e^{ik_{\parallel}z},
    \label{wfxn_LG}
\end{equation}
\end{widetext}
where $\textrm{L}_{p}^{|\ell|}$ are the Laguerre polynomials, with $\ell$ and $p$ being the azimuthal and radial quantum numbers, respectively, and $w_0$ is the $z$-independent beam waist. Fig. \ref{F2}(c) displays two different LG modes with finite beam waists $w_{0}$. 

Similarly, the HG family of transversely localized wave functions are solutions to the paraxial wave equation in the Cartesian coordinate system. Since the nondiffracting LG and HG states each comprise a complete orthonormal basis, any LG (HG) state can be synthesized from the appropriate coherent superposition of HG (LG) states \cite{PhysRevLett.109.084801}. Unlike the LG transverse states, the HG states lack a well-defined azimuthal phase, and owing to the fact that they are not eigenstates ${\hat L}_z$, do not carry a single OAM unit of $\ell$. In the nondiffracting limit, the HG states take the form
\begin{widetext}
\begin{equation}
    \psi_{nm}(x,y,z) = \frac{2^{-\frac{n+m}{2}}}{w_0\sqrt{L}}\sqrt{\frac{2}{\pi n! m!}}  \, H_{n} \bigg( \frac{x\sqrt{2}}{w_{0}} \bigg) H_{m} \bigg( \frac{y\sqrt{2}}{w_{0}} \bigg)e^{-\frac{x^2 + y^2}{w_0^{2}}}e^{ik_{\parallel}z},
    \label{wfxn_HG}
\end{equation}
\end{widetext}
and are labeled by the indices $n$ and $m$, corresponding to the order of the $x$- and $y$-dependent Hermite polynomials, $H_{n}$ and $H_{m}$, respectively. Fig. \ref{F2}(d) displays the first-order $x$- and $y$-oriented HG modes with beam waists $w_0$, in contrast to the corresponding delocalized $\Psi_{k_x}(x,y)$ and $\Psi_{k_y}(x,y)$ plane waves displayed in Fig. \ref{F2}(a). 

Mirroring applications of optical OAM states, free electrons with quantized transverse degrees of freedom have recently been recognized as potential carriers of quantum information, specifically as free electron OAM qubits. Realization of such OAM qubits has been made possible via holographic masks and spiral phase plates \cite{verbeeck2010production, uchida2010generation, mcmorran2011electron, PhysRevLett.114.034801} or through tailored light sources \cite{vanacore2019ultrafast, Madan2022-dr}. Stemming from the separability of the electron wave function following condition (ii), the orthogonal transverse degrees of freedom can be used as orthonormal basis states $|0\rangle$ and $|1\rangle$ on the Bloch sphere (Fig. \ref{F2}(e)). Known as the horizontal and vertical basis states, respectively, linear combinations of $|0\rangle$ and $|1\rangle$ assemble the remaining antipodal points $| L / R \rangle = (1 / \sqrt{2})\big[ |0\rangle \pm i|1\rangle \big]$ and $| \pm \rangle = (1 / \sqrt{2})\big[ |0\rangle \pm |1\rangle \big]$. One example are the first order LG states $\Psi_{\pm 1 0}(\rho, \phi)=\langle x_1  x_2| L/R \rangle$ which span the truncated $\ell = \pm 1$, two dimensional Hilbert space \cite{PhysRevLett.109.084801, Loffler2022-lh, L_ffler_2023}, and lie on the north and south poles of the Bloch sphere. Owing to the fact that $\Psi_{\pm 1 0}(\rho, \phi)$ can be expressed as linear combinations of the first order HG states \cite{PhysRevA.45.8185}, $\Psi_{1 0}(x, y) = \langle x_1 x_2|0\rangle$ and $\Psi_{0 1}(x, y) = \langle x_1 x_2|1\rangle$ are the wave functions associated with the vertical and horizontal basis states, respectively. At the center of the Bloch sphere lies the Gaussian mode $\Psi_{00}(x,y)=\Psi_{00}(\rho, \phi) = \langle x_1  x_2| G \rangle=\sqrt{2/\pi w_0^2}e^{-(x^2+y^2)/w_0^2}=\sqrt{2/\pi w_0^2}e^{-\rho^2/w_0^2}$. More generally, higher order electron vortex states with topological charge of $\pm \ell$ can be used to construct Bloch spheres for $\ell$ values other than unity \cite{Yao2011-eq}. Following the work of Ref. \cite{bourgeois2023optical}, under the appropriate limits, certain superpositions of electron plane wave states can likewise be mapped onto points on the Bloch sphere (Fig. \ref{F2}(e)). In this scenario, the horizontal and vertical plane wave wave functions $\Psi_{k_x}(x,y) = \langle x_1  x_2| 0 \rangle$ and $\Psi_{k_y}(x,y) = \langle x_1  x_2| 1 \rangle$ can be used to construct the north and south antipodal point wave functions $\Psi_{k_xk_y}^{\pm \pi / 2}(x,y)$. Specifically, $\Psi_{k_xk_y}^{\pm \pi / 2}(x,y)=\langle x_1  x_2| L/R \rangle  = (1 / \sqrt{2})\big[\Psi_{k_x}(x,y) + \Psi_{k_y}(x,y)e^{i\chi}\big]$ with $\chi = \pm \pi /2$. At the center of the Bloch sphere is located the electron plane wave $\Psi_{k_\parallel}(x,y)=\langle x_1  x_2| G \rangle =1/L$ with spatially uniform transverse phase and purely longitudinal wave vector $\mathbf{k} = k_\parallel \hat{\mathbf{z}}.$ As will be shown in the following section, transitions between OAM or linear momentum electron states residing on the Bloch sphere produce transition current densities with unique vector and phase profiles. 

\section{Transition current density}
\label{sec_Jfi}
\begin{figure*}
    \centering   
\includegraphics{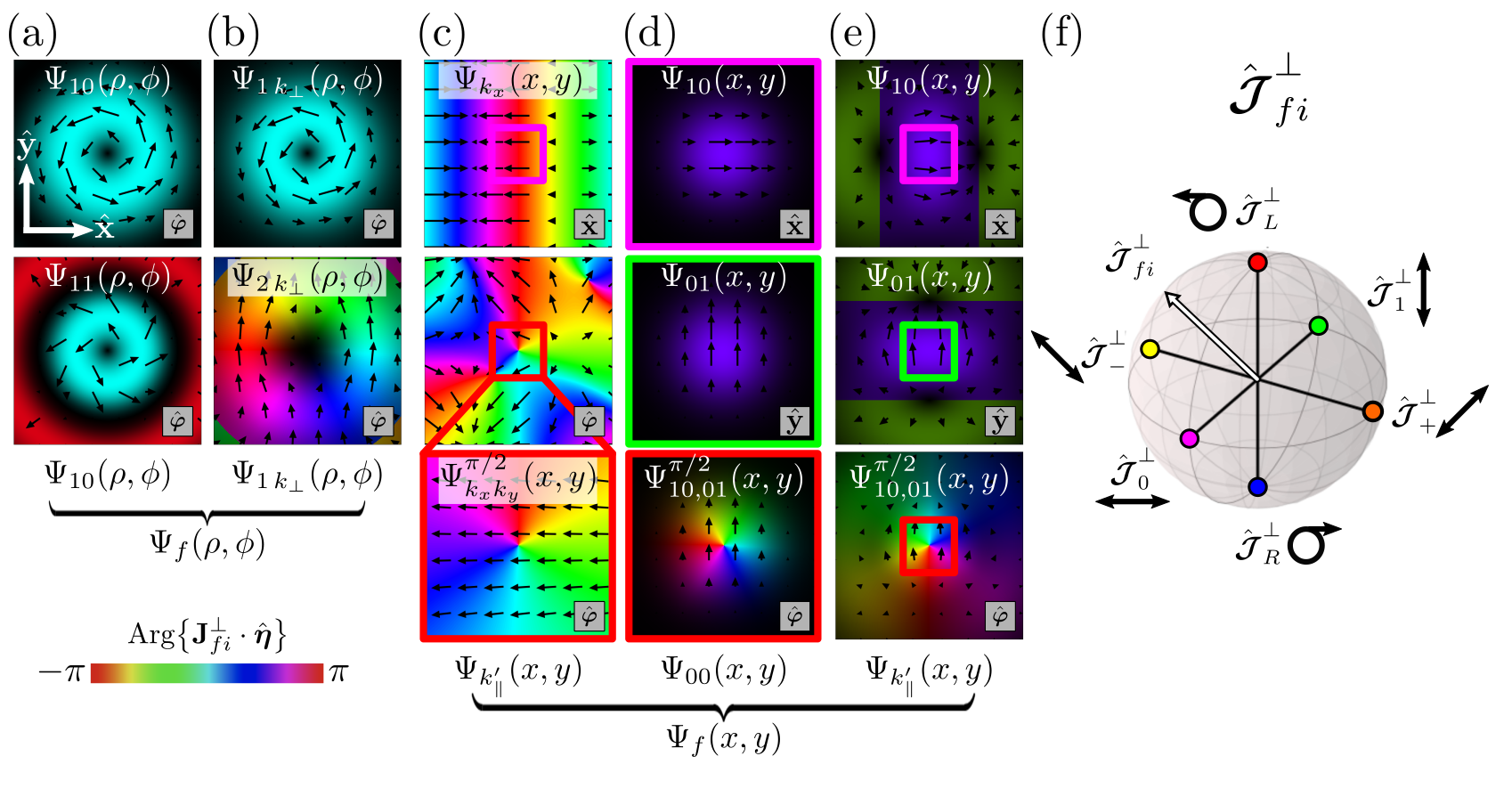}
    \caption{Transition current density vector fields $\mathbf{J}_{fi}({\bf x})$ in the $z=0$ plane. The arrows within each panel show the direction and strength of the transverse component $\mathbf{J}^\perp_{fi}({\bf x})$, while the underlying color map indicates the phase ($\mathbf{J}_{fi}^{\perp}({\bf x}) \cdot \hat{\boldsymbol{\eta}}$) and intensity ($|\mathbf{J}_{fi}^{\perp}({\bf x})|$). $\hat{\boldsymbol{\eta}}$ is displayed in the bottom right hand corner of each panel. The initial transverse wave functions for the transition current density vector fields are labeled in the top-center of each panel, while the final state is shown at the bottom of each column. (a) Transition current densities involving LG $\Psi_{\ell p}(\rho, \phi)$ initial and final states. The top panel shows the case of $\Psi_{10}(\rho, \phi) \to \Psi_{10}(\rho, \phi)$ with no transverse transition; the $\Psi_{11}(\rho, \phi) \to \Psi_{10}(\rho, \phi)$ transition in the lower panel involves a transition in radial, but not OAM, quantum numbers. (b) Transitions between Bessel states $\Psi_{\ell k_\perp}(\rho, \phi)$. The top panel shows the transition current density for $\Psi_{1 k_{\perp}}(\rho, \phi) \to \Psi_{1 k_{\perp}}(\rho, \phi)$ with no transverse state transition, while the lower panel considers the transition current density for $\Psi_{1 k_{\perp}}(\rho, \phi) \to \Psi_{2 k_{\perp}}(\rho, \phi)$ with an OAM change of $\Delta \ell = 1$. (c) Incoming electron plane wave states transitioning to the pin hole state $\Psi_{k^{'}_{\parallel}}(x,y)$. The initial state is characterized by $\mathbf{q}_\perp = k_x \hat{\mathbf{x}}$ in the top panel, while the lower panels involve the incident superposition plane wave state $\Psi^{\pi/2}_{k_x k_y}(x,y)$. The lower panel is a magnified view of the middle panel in the vicinity of the origin. (d) Transition current densities for first-order HG and LG wave functions transitioning to the Gaussian state. (e) Transition current densities for transitions from selected HG and LG states to the pin hole state. (f) $\hat{\boldsymbol{\mathcal{J}}}_{fi}^{\perp}$ for transitions between free electron OAM or linear momentum states which create transition current vector fields resembling the polarization vectors of light on the Poincar\'{e} sphere. The antipodal points $\big\{ \hat{\boldsymbol{\mathcal{J}}}_{0}^{\perp}, \hat{\boldsymbol{\mathcal{J}}}_{1}^{\perp}\big\}$, $\big\{ \hat{\boldsymbol{\mathcal{J}}}_{+}^{\perp}, \hat{\boldsymbol{\mathcal{J}}}_{-}^{\perp}\big\},$ and $\big\{ \hat{\boldsymbol{\mathcal{J}}}_{L}^{\perp}, \hat{\boldsymbol{\mathcal{J}}}_{R}^{\perp}\big\}$ are color coded to match the box colors outlining the panels in columns (c) through (e), as well as the electron state transition arrows in Fig. \ref{F2}(e).}
    \label{F3}
\end{figure*}
Here the transition current density $\mathbf{J}_{fi}(\mathbf{x})$ defined in Eq. \eqref{j_fi} is examined for selected transitions between electron states introduced in the previous section. This section begins with a summary of general properties of $\mathbf{J}_{fi}(\mathbf{x})$ that arise from the restrictions imposed on the forms of the wave functions, then highlights several general and specific forms of the transition current density for transitions involving delocalized and localized electron states. A note regarding the arguments of the transition current density: recall that the origin of time dependence in $\mathbf{J}_{fi}(\mathbf{x},t) = \mathbf{J}_{fi}(\mathbf{x}) e^{i\varepsilon_{fi}t}$ as seen in Eq. (\ref{matrix_elements_V}) is a consequence of condition (i) imposed upon the electron wave functions introduced in the first paragraph of Sec. \ref{electron_states_section}. Therefore, as a matter of convenience and unless stated otherwise, we work with the time-independent version of the transition current density. In addition, when discussing the transition current density for Bessel, LG, and HG free electron states, the subscripts of $\mathbf{J}_{fi}(\mathbf{x})$ refer explicitly to the final and initial transverse electron states.  Panels (a) and (b) of Fig. (\ref{F3}) present LG and Bessel beam transitions, respectively, panel (c) involves transitions between plane wave states, panels (d) and (e) showcase the transition current density using focused first order HG states.

Stemming from conditions (i) and (ii) required of the electron wave functions as stipulated in Sec. \ref{electron_states_section}, the transition current density given by Eq. (\ref{j_fi}) can be re-expressed as the sum of longitudinal ($\parallel$) and transverse ($\perp$) contributions, defined relative to the TEM axis, according to 
\begin{equation}
\begin{split}
    \mathbf{J}_{fi}(\mathbf{x}) &=\mathbf{J}_{fi}(\mathbf{x}_{\perp})e^{iq_{\parallel}z}\\
    &= \big[ \mathbf{J}_{fi}^{\perp}(\mathbf{x}_{\perp}) + J_{fi}^{\parallel}(\mathbf{x}_{\perp})\hat{\bf z}\big ] e^{iq_{\parallel}z}.
    \end{split}
\end{equation}
Explicitly, these components are
\begin{equation}
    \begin{aligned}
        \mathbf{J}_{fi}^{\perp}(\mathbf{x}_{\perp}) & =  \frac{i\hbar e}{2mL}\big[ \Psi_f^{*} (\mathbf{x}_{\perp})\nabla_{\perp}\Psi_i(\mathbf{x}_{\perp}) - \Psi_i(\mathbf{x}_{\perp})\nabla_{\perp} \Psi_f^{*}(\mathbf{x}_{\perp}) \big] \\
        J_{fi}^{\parallel}(\mathbf{x}_{\perp}) & = -\frac{\hbar e}{2mL}Q_{\parallel}\Psi_f^{*} (\mathbf{x}_{\perp}) \Psi_i (\mathbf{x}_{\perp}),
    \end{aligned}
    \label{J_components}
\end{equation}
where $Q_{\parallel} = k_{\parallel} + k_{\parallel}'$. It can be seen from the perpendicular component of Eq. (\ref{J_components}), that interchange of the initial and final transverse states is equivalent to conjugation of the reciprocal scattering process, i.e., $\mathbf{J}_{if}(\mathbf{x}) = \mathbf{J}^{*}_{fi}(\mathbf{x})$.

\subsection{Plane wave states}\label{plane_wave_Jfi_subsec}
First we consider transitions between individual incoming $\psi_{{\bf k}_i}(x,y,z)$ and outgoing $\psi_{{\bf k}_f}(x,y,z)$ plane wave states of the probe, defined in Eq. (\ref{wfxn_epw}), such that $\hbar {\bf q}= \hbar ({\bf k}_i-{\bf k}_f)$ is the momentum recoil associated with the transition. The transition current density associated with this case is 
\begin{equation}
     \mathbf{J}_{{{\bf k}_f}{{\bf k}_i}}(x,y,z) = -\frac{e\hbar}{2mL^3} (2\mathbf{k}_i - \mathbf{q}) e^{i \mathbf{q}\cdot \mathbf{x}}.
     \label{J_fi_pw}
\end{equation}
Inspection of Eq. (\ref{J_fi_pw}) reveals that the transverse component of the transition current density is directly proportional to the transverse recoil momentum, i.e., $\mathbf{J}_{{{\bf k}_f}{{\bf k}_i}}^{\perp}(x,y,z) \propto \mathbf{q}_{\perp}$. The phase and vector structure of the transition current density for an initial plane wave state with transverse wave vector $\mathbf{k}_{\perp}= k_x\hat{\mathbf{x}}$ transitioning to the outgoing plane wave state $\psi_{k'_\parallel}(x,y,z)=L^{-3/2}e^{ik'_\parallel z}$ is shown in the first panel of Fig. \ref{F3}(c). This final state is referred to as the pin hole state as it is selected by placement of a pin hole on the TEM axis in the Fourier plane. In Fig. \ref{F3}(c), it is evident that the periodicity of the plane wave states apparent in Fig. \ref{F2}(a) is inherited by the transition current density.

When prepared in a superposition plane wave state $\psi_{k_xk_y}^{\chi}(x,y,z)$ and post-selected in the diffraction plane for the pin hole state, the resulting transition current density of the probe is
\begin{widetext}
\begin{equation}
    \begin{aligned}
        \mathbf{J}_{k_\parallel',k_xk_y}^\chi(x,y,z) & = -\frac{e\hbar }{2\sqrt{2}mL^{3}}\big[ k_x e^{ik_xx}\hat{\mathbf{x}} + k_y e^{i\chi} e^{ik_yy}\hat{\mathbf{y}}  + Q_{\parallel} \big( e^{ik_xx} + e^{i\chi} e^{ik_yy} \big)\hat{\mathbf{z}}\big]e^{iq_\parallel z}.
    \end{aligned}
    \label{J_fi_pw_super}
\end{equation}
\end{widetext}
The phase and vector information of this transition current density are displayed in the final two panels of Fig. \ref{F3}(c) for $\chi = \pi / 2$ at two different magnification levels. Owing to the spatial dependence of the wave functions, the $\hat{\mathbf{x}}, \hat{\mathbf{y}}$-components of Eq. (\ref{J_fi_pw_super}) are functionally dependent upon $x$ and $y$. It was shown in Ref. \cite{bourgeois2023optical} when working in the dipole scattering regime, whereby $\mathbf{q}_{\perp}$ is such that $|\mathbf{q}_{\perp}| d \ll 1$, with $d$ the transverse length scale of the target, $\mathbf{J}_{fi}^{\perp}(\mathbf{x}_{\perp})$ becomes approximately independent of position in the vicinity of the target. Under these circumstances when the perpendicular components of the transition current density become independent of position, ${\mathbf{J}}_{fi}^{\perp}(\mathbf{x}_{\perp}) \to {\boldsymbol{\mathcal{J}}}_{fi}^{\perp}$. After normalization, $\hat{\boldsymbol{\mathcal{J}}}_{fi}^{\perp}={\boldsymbol{\mathcal{J}}}_{fi}^{\perp}/|{\boldsymbol{\mathcal{J}}}_{fi}^{\perp}|$ can thus be mapped onto the Poincar\'{e} sphere (Fig. \ref{F3}(f)) \cite{bourgeois2023optical}. In this limit, for $\chi = \pi/2 $, the transverse component of Eq. (\ref{J_fi_pw_super}) becomes circularly polarized such that it mimics the polarization of a left-handed circularly polarized photon, evident in the final panel of Fig. \ref{F3}(c) boxed in red. All together, when $|\mathbf{q}_{\perp}| d \ll 1$, the transition current densities given by Eqs. (\ref{J_fi_pw}) and (\ref{J_fi_pw_super}) can be tailored through  appropriate pre- and post-selection to resemble any polarization state of light on the Poincar\'{e} sphere as shown in Fig. \ref{F3}(f).

\subsection{Bessel and Laguerre-Gauss states}
By virtue of the ability to prepare free electrons in defocused vortex (Bessel) or focused vortex (LG) states, the corresponding transition current density can exhibit both radial and azimuthal transverse vector components. The transition current density associated with transitions between Bessel beam states $\psi_{\ell k_\perp}(\rho,\phi,z)$ and $\psi_{\ell' k'_\perp}(\rho,\phi,z)$ is \cite{bourgeois2023optical}
\begin{widetext}
\begin{equation}
    \begin{aligned}
        \mathbf{J}_{\ell' k_{\perp}', \ell k_{\perp}}(\rho,\phi ,z) & = \frac{i\hbar e}{2m} \Psi_{\ell k_{\perp}}(\rho,\phi)  \Psi_{\ell' k_{\perp}'}^{*}(\rho,\phi)\bigg[ \bigg\{ \frac{k_{\perp}}{2}  J_{|\ell|}(k_{\perp} \rho)^{-1}\big[ J_{|\ell| - 1}(k_{\perp} \rho) - J_{|\ell| + 1}(k_{\perp} \rho)\big] \\ 
        & \ \ \ - \frac{k_{\perp}'}{2} J_{|\ell'|}^{*}(k_{\perp}' \rho)^{-1}\big[ J_{|\ell'| - 1}^{*}(k_{\perp}' \rho) - J_{|\ell'| + 1}^{*}(k_{\perp}' \rho)\big] \bigg\} \hat{\bm{\rho}} + \frac{i}{\rho} (\ell +\ell') \hat{\bm{\phi }}  + iQ_{\parallel}\hat{\mathbf{z}} \bigg]e^{i q_\parallel z},
    \end{aligned}
    \label{J_fi_Bes}
\end{equation}
\end{widetext}
for arbitrary $\ell, k_{\perp}$ and $\ell', k_{\perp}'$. The transition current density for $\ell' = \ell = 1$ and $k_{\perp}' = k_{\perp}$, with $\Delta \ell = 0$ is displayed in the first panel of Fig. \ref{F3}(b), showcasing the expected azimuthal character inherent to twisted electron beams. Bessel beam states can also lead to unique phase and vector profiles as seen in the lower panel of Fig. \ref{F3}(b), wherein $\Delta \ell = 1$ resulting in a transition current density with a phase-structure similar to that of circularly polarized light, as well as to the wave functions shown in Fig. \ref{F2}(b).

Additionally, the focused LG states in Eq. (\ref{wfxn_LG}) can also produce transition current densities with well-defined units of OAM transferred. The transition current density for an initial LG electron state $\psi_{\ell p}(\rho, \phi ,z)$ transitioning to a final state $\psi_{\ell' p'}(\rho,\phi ,z)$ is given by
\begin{widetext}
\begin{equation}
    \begin{aligned}
        \mathbf{J}_{\ell' p', \ell p}(\rho, \phi ,z)
        & = \frac{i\hbar e}{2 m} \Psi_{\ell p}(\rho,\phi)  \Psi_{\ell' p'}^{*}(\rho,\phi) \bigg[ \frac{1}{\rho} \bigg\{ \big( |\ell| - |\ell'|\big) \\
        & \ \ \ - \frac{4\rho^2}{w_0^2} \bigg( \textrm{L}_{p}^{|\ell|}\bigg( \frac{2\rho^2}{w^{2}_0} \bigg) \textrm{L}_{p'}^{|\ell'|*} \bigg( \frac{2\rho^2}{w^{2}_0} \bigg) \bigg)^{-1}\bigg( \textrm{L}_{p'}^{|\ell'|*} \bigg( \frac{2\rho^2}{w^{2}_0} \bigg) \textrm{L}_{p}^{|\ell|+1} \bigg( \frac{2\rho^2}{w^{2}_0} \bigg) - \textrm{L}_{p}^{|\ell|}\bigg( \frac{2\rho^2}{w^{2}_0} \bigg) \textrm{L}_{p'}^{|\ell'|+1,*} \bigg( \frac{2\rho^2}{w^{2}_0} \bigg) \bigg) \bigg\} \hat{\boldsymbol{\rho}}\\
        & \ \ \ +\frac{i}{\rho}(\ell + \ell')\hat{\boldsymbol{\phi }} + iQ_{\parallel}\hat{\mathbf{z}} \bigg]e^{i q_\parallel z}
    \end{aligned}
    \label{J_fi_LG}
\end{equation}
\end{widetext}
valid for arbitrary $\ell, p$ and $\ell', p'$. Stemming from the commonalities in the underlying wave functions, it is unsurprising that Eq. (\ref{J_fi_Bes}) and Eq. (\ref{J_fi_LG}) look similar. This resemblance is most obvious when considering events wherein no transverse transition occurs, as seen when comparing the first panels of Fig. \ref{F3}(a) and Fig. \ref{F3}(b), for transitions between LG and Bessel states, respectively. Here, shown in the top panel of Fig. \ref{F3}(a), is Eq. (\ref{J_fi_LG}) for $p' = p = 0$ and $\ell' = \ell = 1$, which highlights the azimuthal vector component of the LG transition current density and is similar to the corresponding Bessel beam case displayed in the top panel of Fig. \ref{F3}(b) for $1k_\perp$ transitioning to $1k_\perp$. The bottom panel of Fig. \ref{F3}(a) shows the transition current density for $\Delta p = 1$. For inelastic scattering events where the electron does not transfer OAM to the target, $\ell' = \ell$ and $p' = p$, and Eq. (\ref{J_fi_LG}) reduces to 
\begin{equation}
    \begin{aligned}
        \mathbf{J}_{\ell'=\ell, p'=p}(\rho, \phi ,z) = -\frac{\hbar e}{2 m L} \big| \Psi_{\ell p}(\rho,\phi ) \big|^2 \Big[ \frac{2\ell }{\rho}\hat{\boldsymbol{\phi }} + Q_{\parallel}\hat{\mathbf{z}} \Big]e^{i q_\parallel z},
    \end{aligned}
    \label{J_fi_eqLG}
\end{equation}
which is the focused vortex beam form of the transition current density of a free electron moving along $\mathbf{\hat{v}} = \mathbf{\hat{z}}$ \cite{bliokh2017theory, RevModPhys.89.035004}. The azimuthal component of Eq. (\ref{J_fi_eqLG}) is responsible for the spiraling behavior of the electron current typical for vortex beams, and upon spatial integration over $(\rho, \phi)$ yields an electron current with an axial inertial OAM equal to $\ell\hbar$ \cite{bliokh2017theory}.

\subsection{Hermite-Gauss states}
Transitions between transverse HG states that are naturally expressed in the Cartesian basis provide transition current densities with any desired $ \hat{\mathbf{x}}, \hat{\mathbf{y}}$-vectorial structure. Following the approach of Ref. \cite{bourgeois2023optical}, a general form for $\mathbf{J}_{n'm', nm}(x,y,z)$ involving arbitrary HG states can be derived from Eq. (\ref{j_fi}). The transition current density for $\psi_{nm}(x,y,z)$ transitioning to $\psi_{n'm'}(x,y,z)$ is
\begin{widetext}
\begin{equation}
    \begin{aligned}
        \mathbf{J}_{n'm', nm}(x,y,z) & = \frac{i \hbar e }{2mw_0 L} \bigg[ \big\{ \sqrt{n}\Psi_{n-1,m}(x,y)\Psi_{n'm'}^{*}(x,y) - \sqrt{n+1}\Psi_{n+1,m}(x,y)\Psi_{n'm'}^{*}(x,y) \\
        & \ \ \ - \sqrt{n'}\Psi_{nm}(x,y)\Psi_{n'-1,m'}^{*}(x,y) 
        + \sqrt{n'+1}\Psi_{nm}(x,y)\Psi_{n'+1,m'}^{*}(x,y) \big\}\hat{\mathbf{x}} \\
        & \ \ \ + \big\{ \sqrt{m}\Psi_{n,m-1}(x,y)\Psi_{n'm'}^{*}(x,y)-\sqrt{m+1}\Psi_{n,m+1}(x,y)\Psi_{n'm'}^{*}(x,y) \\
        & \ \ \ -\sqrt{m'}\Psi_{nm}(x,y)\Psi_{n',m'-1}^{*} (x,y)+\sqrt{m'+1}\Psi_{nm}(x,y)\Psi_{n',m'+1}^{*}(x,y) \big\}\hat{\mathbf{y}} \\
        & \ \ \ + i w_0 Q_{\parallel}\Psi_{nm}(x,y)\Psi_{n'm'}^{*}(x,y) \hat{\mathbf{z}} \bigg] e^{i q_\parallel z},
    \end{aligned}
    \label{J_HG}
\end{equation}
\end{widetext}
for arbitrary $n,m, n', m'$. As seen in the form of the perpendicular component of Eq. (\ref{J_HG}), careful choice regarding the pre- and post-selection of the initial/final HG transverse electron states can be intuited to yield a transition current density with non-trivial $ \hat{\mathbf{x}}, \hat{\mathbf{y}}$-vectorial behavior in the plane orthogonal to the direction of propagation. Working in the $\ell = \pm 1$ OAM Hilbert space, $\mathbf{J}_{n'm', nm}(x,y,z)$ is displayed in the columns of Fig. \ref{F3}(d) for first order electron wave functions 
$\psi_{10}(x,y,z)$ and $\psi_{01}(x,y,z)$, and their linear combination $\psi_{10, 01}^{\chi=\pi/2}(x,y,z) = (1 / \sqrt{2L}) \big[ \psi_{10}(x,y,z) +\psi_{01}(x,y,z)e^{i\pi/2}\big]$, respectively, all transitioning to the final Gaussian wave function $\psi_{00}(x,y,z) = (1/\sqrt{L})\Psi_{00}(x,y)e^{ik_{\parallel}'z}$. In the zero width limit whereby $w_0 \rightarrow 0$, for the first order transitions discussed above, the transverse components of the transition current density become spatially independent and $\mathbf{J}^\perp_{fi}({\bf x})\to{\boldsymbol{\mathcal{J}}}^\perp_{fi}$. Under these constraints, the transition current density unit vector ${\boldsymbol{\mathcal{J}}}^\perp_{fi}$ can thus imitate the polarization vector $\hat{\boldsymbol{\epsilon}}$ of free space light, as illustrated on the electron analog of the Poincar\'{e}
sphere presented in Fig. \ref{F3}(f). When no transverse transition of the probe occurs, i.e., $n'=n$ and $m'=m$ in Eq. (\ref{J_HG}), the transition current density reduces to
\begin{equation}
    \mathbf{J}_{n'=n,m'=m}(x,y,z) = -\frac{e\hbar }{2mL} Q_{\parallel} \big| \Psi_{n m}(x,y) \big|^2 e^{i q_\parallel z}\hat{\mathbf{z}},
    \label{J_HGneqm}
\end{equation}
which is oriented parallel to the TEM axis regardless of the values for $n$ and $m$. In the zero width limit, Eq. (\ref{J_HGneqm}) reduces to the classical current of a point electron source \cite{de2010optical} when $n=m=0$. Overall, the purely longitudinal behavior of Eq. (\ref{J_HGneqm}) differs from the vortex electron beam case in Eq. (\ref{J_fi_eqLG}) wherein the current density has a $\hat{\boldsymbol{\phi }}$-component perpendicular to the electron's direction of motion. 

Alternatively, when the $nm$ HG state transitions to the forward directed pin hole state rather than a transversely focused Gaussian state, the transition current density becomes
\begin{widetext}
\begin{equation}
    \begin{aligned}
        \mathbf{J}_{k_\parallel',nm}(x,y,z) & = \frac{i\hbar e}{2mw_0L^2}\bigg[ \big\{ \sqrt{n}\Psi_{n-1,m}(x,y) - \sqrt{n+1}\Psi_{n+1,m}(x,y) \big\}\hat{\mathbf{x}} \\
        &\ \ \  + \big\{ \sqrt{m}\Psi_{n,m-1}(x,y) - \sqrt{m+1}\Psi_{n,m+1}(x,y) \big\}\hat{\mathbf{y}} \\
        &\ \ \  +iw_0Q_{\parallel} \Psi_{n,m}(x,y) \hat{\mathbf{z}} \bigg] e^{iq_\parallel z}.
    \end{aligned}
    \label{J_HG_pw_final_pinhol}
\end{equation} 
\end{widetext}
Eq. (\ref{J_HG_pw_final_pinhol}) is displayed in Fig. \ref{F3}(e) for the same initial HG states as those used in Fig. \ref{F3}(d), transitioning instead to the pin hole state $\psi_{k'_\parallel}(x,y,z)$. For the focused superposition state $\psi_{10,01}^{\chi}(x,y,z) = (1 / \sqrt{2L}) \big[ \psi_{10}(x,y,z) +\psi_{01}(x,y,z)e^{i\chi}\big]$ transitioning to the pin hole state $\psi_{k'_\parallel}(x,y,z)$, the transition current density explicitly has the form
\begin{widetext}
\begin{equation}
    \begin{aligned}
        \mathbf{J}_{k_{\parallel}',10,01}^{\chi}(x,y,z) & = \frac{i\hbar e}{2\sqrt{2}mw_0L^2}\bigg[ \big\{\Psi_{00}(x,y) - \sqrt{2}\Psi_{20}(x,y) - \Psi_{11}(x,y)e^{i\chi} \big\}\hat{\mathbf{x}} \\
        &\ \ \  + \big\{ \big(\Psi_{00}(x,y) - \sqrt{2}\Psi_{02}(x,y) \big)e^{i\chi} - \Psi_{11}(x,y)\big\} \hat{\mathbf{y}} \\
        &\ \ \  + iw_0Q_{\parallel} \big\{ \Psi_{10}(x,y) +\Psi_{01}(x,y)e^{i\chi }\big\}\hat{\mathbf{z}} \bigg] e^{iq_\parallel z}.
    \end{aligned}
    \label{J_fi_hg_sp}
\end{equation} 
\end{widetext}
The difference between Eqs. (\ref{J_fi_pw_super}) and (\ref{J_fi_hg_sp}) originates from the choice of the initial electron wave function being either a superposition of linear momentum or OAM states, respectively. Imposing the same conditions placed upon the wave functions in Eq. (\ref{J_HG}) to obtain a transition current density whose transverse vectorial components are spatially independent, Eqs. (\ref{J_HG_pw_final_pinhol}) and (\ref{J_fi_hg_sp}) can be used to construct transition current densities on the surface of the Poincar\'{e} sphere as presented in Fig. \ref{F3}(f). Therefore, transitions between specific states on the Bloch sphere (Fig. \ref{F2}(e)) produce transition current densities which mimic the polarization structure of free space light and can thus be mapped onto the Poincar\'{e} sphere (Fig. \ref{F3}(f)).

\section{State- and energy-resolved Observables}\label{sec_observables}
Building from Secs. \ref{theory_IES}, \ref{electron_states_section}, and \ref{sec_Jfi}, in this section the EEL, EEG, and DDCS observables between phase-shaped states of the electron probe are derived. Measurements of EEL and EEG processes are discussed first, including the narrow beam width limit common in the low-loss electron scattering regime, before moving on to presentation of the DDCS. A comparison of the EEL and EEG scattering processes, and their relation to the properties of the transition current density under interchange of initial/final electron states is briefly presented. We illustrate that the transversely phase-shaped EEL, EEG, and DDCS observables, reduce to the familiar forms found in the literature under the appropriate limits.

\subsection{Electron energy loss and gain probabilities in the narrow beam width limit}
\label{narrow_beam_section}
In considering low-loss EEL and EEG scattering events in the narrow beam width limit appropriate to the STEM it is customary to work within the nonrecoil approximation, where the change in the energy of the electron is dictated entirely by its momentum change along the axis of propagation \cite{de2010optical}. In this limit, the forward recoil momentum $\hbar q_\parallel \hat{\bf z}$ is small compared to the electron's initial momentum $\hbar{\bf k}_i$ so that its change in energy can be approximated by $\hbar \varepsilon_{if} = \sqrt{(mc^2)^2 + (\hbar c \mathbf{k}_i)^2} - \sqrt{(mc^2)^2 + (\hbar c (\mathbf{k}_i - \mathbf{q}))^2} \approx \hbar \mathbf{v}_i \cdot \mathbf{q}$, where $\hbar \mathbf{k}_i /m = \gamma_i \mathbf{v}_i$ and $\hbar \mathbf{k}_i = \mathbf{p}_i$ for relativistic matter waves. Upon insertion of $\varepsilon_{if} = \mathbf{v}_i \cdot \mathbf{q}$ into the trailing delta function in Eq. (\ref{w_fi_loss}), the state- and frequency-resolved EEL rate becomes
\begin{widetext}
\begin{equation}
    w_{fi}^{\textrm{loss}}(\omega)  = -\frac{8 \pi}{\hbar v_i} \int d\mathbf{x} d\mathbf{x}' \, \textrm{Im}\Big[ \mathbf{J}_{fi}^{*}(\mathbf{x}) \cdot \tensor{\mathbf{G}}(\mathbf{x}, \mathbf{x}', \omega) \cdot \mathbf{J}_{fi}(\mathbf{x}') \Big] \delta(q_{\parallel} - \omega / v_i),
\label{state_energy_w_rate_loss}
\end{equation}
\end{widetext}
where $\tensor{\mathbf{G}}(\mathbf{x}, \mathbf{x}', \omega)$ is the target's electromagnetic Green’s tensor introduced in Sec. \ref{theory_IES}. The state- and energy-resolved EEL probability $P_{fi}^{\textrm{loss}}(\omega) = ( L / v_i ) w_{fi}^{\textrm{loss}} \, (\omega)$ is obtained by integrating Eq. (\ref{state_energy_w_rate_loss}) over the time it takes the probe electron to traverse the path length $L$ as it interacts with the target specimen. From $P_{fi}^{\textrm{loss}}(\omega)$, the  state- and energy-resolved EEL probability  is determined by summing over all possible final electron states with $\sum_{k_\parallel^f}\to+(L/2\pi)\int^\infty_{-\infty} dq_{\parallel}$ and dividing by $\hbar$, resulting in the EEL probability per unit energy
\begin{widetext}
\begin{equation}
    \Gamma_{fi}^{\textrm{loss}}(\omega) = -\frac{4}{\hbar^2} \int d\mathbf{x} d\mathbf{x}'\, \textrm{Im}\Big[\bigg(\frac{L}{v_i}\bigg)\mathbf{J}_{fi}^{*}(\mathbf{x}; q_{\parallel} = \frac{\omega}{ v_i}) \cdot \tensor{\mathbf{G}}(\mathbf{x}, \mathbf{x}', \omega) \cdot \bigg(\frac{L}{v_i}\bigg)\mathbf{J}_{fi}(\mathbf{x}'; q_{\parallel} = \frac{\omega}{v_i}) \Big].
    \label{EEL_probability}
\end{equation}
\end{widetext}
Here $\mathbf{J}_{fi}(\mathbf{x};q_{\parallel} = {\omega}/{v_i})=\mathbf{J}_{fi}(\mathbf{x})|_{q_{\parallel} = {\omega}/{v_i}}=\mathbf{J}_{fi}(\mathbf{x}_{\perp})e^{iq_\parallel z}|_{q_{\parallel} = {\omega}/{v_i}}=\mathbf{J}_{fi}(\mathbf{x}_{\perp})e^{i(\omega/v_i)z}$ makes explicit the locking of the longitudinal recoil wave number $q_\parallel$ to $\omega/v_i$ imposed by the nonrecoil approximation in Eqs. (\ref{state_energy_w_rate_loss}) and (\ref{EEL_probability}). Due to the frequent appearance of the $L/v_i$ factor here and in the following equations, we define a new transition current as $\mathbf{J}_{fi}(\mathbf{x}, \omega=q_{\parallel}{v_i}) \equiv(L / v_i) \mathbf{J}_{fi}(\mathbf{x};q_{\parallel} = {\omega}/{v_i})$ with dimensions of charge flux per unit frequency. For clarity, we also abandon the general notation $\mathbf{x} = (\mathbf{x}_\perp,z)$ in favor of $\mathbf{x} = (\mathbf{R},z)$ as is common in the low-loss STEM EEL and EEG literature.

Using the current $\mathbf{J}_{fi}(\mathbf{x}, \omega=q_{\parallel}{v_i})$, the EEL probability can also be cast in terms of the target's transition electric field $\mathbf{E}_{fi}(\mathbf{x}, \omega=q_{\parallel}{v_i}) = -4\pi i \omega \int d\mathbf{x}' \, \tensor{\mathbf{G}}(\mathbf{x}, \mathbf{x}', \omega) \cdot \mathbf{J}_{fi}(\mathbf{x}', \omega=q_{\parallel}{v_i})$ resolved in frequency. $\mathbf{E}_{fi}$ represents the electric field produced by the target in response to stimulation by the probing electron. Expressed in terms of this field, the EEL probability per unit energy becomes
\begin{widetext}
\begin{equation}
\Gamma_{fi}^{\textrm{loss}}(\omega) = -\frac{1}{\pi \hbar^2 \omega } \textrm{Re}\bigg[ \int d\mathbf{x} \,\mathbf{J}_{fi}^{*}(\mathbf{x}, \omega=q_{\parallel}{v_i}) \cdot \mathbf{E}_{fi}(\mathbf{x}, \omega=q_{\parallel}{v_i}) \bigg],
    \label{EEL_probability_induc_field}
\end{equation}
\end{widetext}
which recovers the classical relationship \cite{PhysRev.106.874, de2010optical} between $\Gamma_{fi}^{\textrm{loss}}(\omega)$ and the work performed by the electron against its own induced field. When the electron beam waist $w_{0}$ is negligible compared to the length scale over which the response field changes, $\mathbf{E}_{fi}$ can be taken as constant over the spatial domain where the current density is appreciable and approximated by its value $\mathbf{E}_{fi}(\mathbf{R}, z, \omega=q_{\parallel}{v_i}) \approx \mathbf{E}_{fi}(\mathbf{R}_0, z, \omega=q_{\parallel}{v_i})$ at the impact parameter ${\bf R}_0$. In this narrow beam width limit, Eq. (\ref{EEL_probability_induc_field}) becomes
\begin{widetext}
\begin{equation}
    \begin{aligned}
        \Gamma_{fi}^{\textrm{loss}}(\mathbf{R}_0, \omega) & = -\frac{1}{\pi \hbar^2 \omega } \textrm{Re}\bigg[ \int dz \bigg( \int d\mathbf{R} \, \mathbf{J}_{fi}^{*}(\mathbf{R},z ,\omega=q_{\parallel}{v_i}) e^{i\omega z/v_i}\bigg) e^{-i\omega z/v_i} \cdot \mathbf{E}_{fi}(\mathbf{R}_0,z,\omega=q_{\parallel}{v_i}) \bigg] \\
        & = -\frac{1}{\pi \hbar^2 \omega } \textrm{Re}\bigg[\boldsymbol{\mathcal{J}}^{*}_{fi} \cdot \int dz \,\mathbf{E}_{fi}(\mathbf{R}_0,z,\omega=q_{\parallel}{v_i}) e^{-i\omega z/v_i} \bigg],
    \end{aligned}
    \label{EEL_probability_induc_field_ff}
\end{equation}
\end{widetext}
where the transition current $\boldsymbol{\mathcal{J}}_{fi} = \int d\mathbf{R} \, \mathbf{J}_{fi}(\mathbf{R}, z,\omega=q_{\parallel}{v_i})e^{-i\omega z/v_i} = (-{e}/{mv_i})\big[\langle \Psi_f | \hat{\mathbf{p}}_{\perp}|\Psi_i\rangle + (\hbar Q_{\parallel} /2)  \langle\Psi_f | \Psi_i\rangle \hat{\mathbf{z}}\big]$ is $z$- and $\omega$-independent. It has the transverse and longitudinal components $\boldsymbol{\mathcal{J}}_{fi}^{\perp} = -( e/mv_i) \langle \Psi_{f}| \hat{\mathbf{p}}_{\perp} | \Psi_{i}\rangle$ and $\boldsymbol{\mathcal{J}}_{fi}^{\parallel} = -(\hbar e Q_{\parallel}/2mv_i) \langle \Psi_{f} | \Psi_{i} \rangle \hat{\mathbf{z}}$, respectively, which allow the EEL probability in Eq. (\ref{EEL_probability_induc_field_ff}) to be separated into the perpendicular and parallel contributions
\begin{widetext}
\begin{equation}
    \begin{aligned}
        \Gamma_{fi\perp}^{\textrm{loss}}(\mathbf{R}_0, \omega) &  = -\frac{1}{\pi \hbar^2 \omega } \textrm{Re}\bigg[ \boldsymbol{\mathcal{J}}_{fi}^{\perp *} \cdot \int dz \,\mathbf{E}_{fi}(\mathbf{R}_0,z,\omega=q_{\parallel}{v_i}) e^{-i\omega z/v_i} \bigg]\\
        \Gamma_{fi\parallel}^{\textrm{loss}}(\mathbf{R}_0, \omega) &  = -\frac{1}{\pi \hbar^2 \omega } \textrm{Re}\bigg[\boldsymbol{\mathcal{J}}_{fi}^{\parallel *} \cdot \int dz \,\mathbf{E}_{fi}(\mathbf{R}_0,z,\omega=q_{\parallel}{v_i}) e^{-i\omega z/v_i} \bigg].
    \end{aligned}
    \label{EEL_perp_and_par}
\end{equation}
\end{widetext}
Based on the forms of $\boldsymbol{\mathcal{J}}_{fi}^{\perp}$ and $\boldsymbol{\mathcal{J}}_{fi}^{\parallel}$ above, it is evident that $\boldsymbol{\mathcal{J}}_{fi}^{\perp}={\bf 0}$ and $\boldsymbol{\mathcal{J}}_{fi}^{\parallel}=-(\hbar e Q_{\parallel}/2mv_i)\hat{\bf z}$ in the event of no transverse transition (i.e., $\Psi_i=\Psi_f$ and $\Gamma_{fi\perp}^{\textrm{loss}}=0$), while $\boldsymbol{\mathcal{J}}_{fi}^{\perp}\neq{\bf 0}$ and $\boldsymbol{\mathcal{J}}_{fi}^{\parallel} = {\bf 0}$ when a transition occurs in the probe's transverse wave function (i.e., $\Psi_i\neq\Psi_f$ and $\Gamma_{fi\parallel}^{\textrm{loss}} = 0$). In the former case, $\Gamma_{fi\parallel}^{\textrm{loss}}(\mathbf{R}_0, \omega)$ in Eq. (\ref{EEL_perp_and_par}) reduces to the the well known classical form for the EEL probability \cite{de2008probing, hohenester2009electron, de2010optical, garcia2021optical} in the zero width limit $w_0\to0.$ Specifically, $\Gamma_{fi\parallel}^{\textrm{loss}}(\mathbf{R}_0, \omega)\to\Gamma_{\textrm{cl}} (\mathbf{R}_0, \omega) = -(2e/\hbar)^2\int dzdz'\, \textrm{Im}[\hat{\bf z} \cdot\tensor{\mathbf{G}}(\mathbf{R}_0,z; \mathbf{R}_0,z', \omega)\cdot\hat{\bf z} \, e^{-i\omega(z-z')/v_i}]$, which is the EEL probability per unit energy for a uniformly moving classical electron with current density $\mathbf{J}_{\textrm{cl}}(\mathbf{x}, \omega) = -e\delta(\mathbf{R} - \mathbf{R}_0)e^{i\omega z / v_i} \, \hat{\mathbf{z}}$ at impact parameter ${\bf R}_0$.

To construct the laser-stimulated phase-shaped EEG observable in the narrow beam width limit, we again introduce $\varepsilon_{if} =  \mathbf{v}_i \cdot \mathbf{q}$ into the trailing delta function in the frequency-resolved EEG rate presented in Eq. (\ref{w_fi_gain}). As a result, the state- and frequency-resolved EEG rate and scattering probability is
\begin{widetext}
\begin{equation}
    w_{fi}^{\textrm{gain}}(\omega) = \frac{2\pi}{v_i}\bigg( \frac{1}{\hbar \omega_\nu} \bigg)^2  \bigg| \int d\mathbf{x} \, \mathbf{E}_{\nu}^{(+)}(\mathbf{x}) \cdot \mathbf{J}_{fi}(\mathbf{x}) \bigg|^2 \delta(q_{\parallel} + \omega/ v_i)\delta(\omega - \omega_{\nu\nu'})
\end{equation}
\end{widetext}
and $P_{fi}^{\textrm{gain}}(\omega) = \big( L / v_i \big) w_{fi}^{\textrm{gain}} \, (\omega)$. In parallel to loss, the EEG probability per unit energy is determined from $P_{fi}^{\textrm{gain}}(\omega)$ after integrating over final states $\sum_{k^f_\parallel}\to+ (L/2\pi)\int_{-\infty}^{\infty} dq_{\parallel}$ and dividing by $\hbar$, resulting in
\begin{widetext}
\begin{equation}
    \Gamma_{fi}^{\textrm{gain}}(\omega) = \frac{1}{\hbar}\bigg( \frac{1}{\hbar \omega_\nu} \bigg)^2  \bigg| \int d\mathbf{x} \, \mathbf{E}_{\nu}^{(+)}(\mathbf{x}) \cdot \mathbf{J}_{fi}(\mathbf{x}, \omega=-q_{\parallel}v_i) \bigg|^2 \delta(\omega - \omega_{\nu\nu'}).
    \label{gain_prob_gen}
\end{equation}
\end{widetext}
As in the case of loss, if the target's induced electric field varies little over the spatial domain of the probe's transition current density, then $\mathbf{E}_{\nu}^{(+)}(\mathbf{R},z)\approx\mathbf{E}_{\nu}^{(+)}(\mathbf{R}_0,z)$ to lowest order and the state- and energy-resolved EEG probability takes the form
\begin{widetext}
\begin{equation}
    \begin{aligned}
        \Gamma_{fi}^{\textrm{gain}}(\mathbf{R}_0, \omega) & = \frac{1}{\hbar}\bigg( \frac{1}{\hbar \omega_\nu } \bigg)^2  \bigg| \int dz \, \mathbf{E}_{\nu}^{(+)}(\mathbf{R}_0,z)e^{-i\omega z/v_i} \cdot \bigg[ \int d\mathbf{R} \, \mathbf{J}_{fi}(\mathbf{R},z, \omega=-q_{\parallel}v_i)e^{i\omega z/v_i}\bigg] \bigg|^2 \delta(\omega - \omega_{\nu\nu'}) \\
        & = \frac{1}{\hbar}\bigg( \frac{1}{\hbar \omega_\nu } \bigg)^2  \bigg| \int dz \, e^{-i\omega z/v_i} \mathbf{E}_{\nu}^{(+)}(\mathbf{R}_0,z) \cdot \boldsymbol{\mathcal{J}}_{fi} \bigg|^2 \delta(\omega - \omega_{\nu\nu'}), 
    \end{aligned}
    \label{EEG_NBWL_final_form}
\end{equation}
\end{widetext}
with $z$- and $\omega$-independent transition current $\boldsymbol{\mathcal{J}}_{fi}$ defined below Eq. (\ref{EEL_probability_induc_field_ff}). As in the case of EEL, the EEG probability can also be broken into perpendicular and parallel components
\begin{widetext}
\begin{equation}
    \begin{aligned}
        \Gamma_{{fi}\perp}^{\textrm{gain}}(\mathbf{R}_0, \omega) & = \frac{1}{\hbar}\bigg( \frac{1}{\hbar \omega_\nu } \bigg)^2 \big|{\mathcal{J}}_{fi}^{\perp} \big|^2 \bigg| \int dz \, e^{-i\omega z/v_i}\mathbf{E}_{\nu}^{(+)}(\mathbf{R}_0,z) \cdot \hat{\boldsymbol{\mathcal{J}}}_{fi}^{\perp} \bigg|^2 \delta(\omega - \omega_{\nu\nu'})\\
        \Gamma_{fi\parallel}^{\textrm{gain}}(\mathbf{R}_0, \omega) & = \frac{1}{\hbar}\bigg( \frac{1}{\hbar \omega_\nu } \bigg)^2  \big|{\mathcal{J}}_{fi}^{\parallel} \big|^2 \bigg| \int dz \, e^{-i\omega z/v_i}\mathbf{E}_{\nu}^{(+)}(\mathbf{R}_0,z) \cdot \hat{\boldsymbol{\mathcal{J}}}_{fi}^{\parallel} \bigg|^2 \delta(\omega - \omega_{\nu\nu'}),
    \end{aligned}
    \label{EEG_perp_and_par}
\end{equation}
\end{widetext}
where ${\boldsymbol{\mathcal{J}}}_{fi}=\hat{\boldsymbol{\mathcal{J}}}_{fi}|{\boldsymbol{\mathcal{J}}}_{fi}|=\hat{\boldsymbol{\mathcal{J}}}_{fi}{{\mathcal{J}}}_{fi}$. Similarly, when the transverse wave functions $\Psi_i=\Psi_f$, $\Gamma_{fi}^{\textrm{gain}}(\mathbf{R}_0, \omega)=\Gamma_{fi\parallel}^{\textrm{gain}}(\mathbf{R}_0, \omega)$ and recovers the conventional EEG probability \cite{de2008electron, de2010optical, asenjo2013plasmon, liu2019continuous, bourgeois2022polarization} in the narrow beam limit $w_0\to0$. Through combination of optical polarization and pre- and post-selection of the probe's transverse phase profile to define its polarization, cross-polarized measurements in the STEM can be leveraged to directly interrogate optically-excited target mode symmetries in 3D  with nanoscale spatial resolution \cite{bourgeois2022polarization}.

The above expressions for EEL and EEG probabilities involve interrogation of the target's induced response field by the transition current densities $\mathbf{J}_{fi}(\mathbf{x}, \omega=+q_{\parallel}v_i)$ and $\mathbf{J}_{fi}(\mathbf{x}, \omega=-q_{\parallel}v_i)$ of the probe. It is natural to consider the relationship between these currents upon interchanging initial and final probe states in both the transverse and axial directions. The reciprocal behavior of pre- and post-selection of the probe's transverse states, i.e., $\mathbf{J}_{fi}^{\perp  }(\mathbf{R})=\mathbf{J}_{if}^{\perp *}(\mathbf{R})$, were discussed previously in Sec. \ref{sec_Jfi}. Additionally, along the TEM axis, the longitudinal component $J_{fi}^{\parallel}(\mathbf{R})=J_{if}^{\parallel *}(\mathbf{R})$ expresses the relationship between anti-Stokes (EEG) and Stokes (EEL) scattering processes at the level of the transition current density. Taken together, 
\begin{equation}
    \begin{aligned}
        \mathbf{J}_{fi}(\mathbf{x}, \omega=+q_{\parallel}{v_i})
        &=\Big(\frac{L}{ v_i}\Big) \big[ \mathbf{J}_{if}^{\perp *}(\mathbf{R}) + J_{if}^{\parallel *}(\mathbf{R})\hat{\bf z}\big ] e^{i(\omega/v_i)z}\\
        &=\Big(\frac{L}{ v_i}\Big) \Big(\big[ \mathbf{J}_{if}^{\perp }(\mathbf{R}) + J_{if}^{\parallel }(\mathbf{R})\hat{\bf z}\big ] e^{-i(\omega/v_i)z}\Big)^*\\
        &=\mathbf{J}_{if}^*(\mathbf{x}, \omega=-q_{\parallel}{v_i}).
    \end{aligned}
\end{equation}
These symmetries of the transition current density under interchange of initial and final states, and their effect on the observables, is detailed further in the Appendix.

\subsection{Double differential inelastic scattering cross section in the wide field limit}
\label{DDCS_section}
When dealing with plane wave electron states, the scattering cross section is a common observable of interest. It is attained from the EEL transition rate $w^{\textrm{loss}}_{fi}(\omega)$ in Eq. (\ref{w_fi_loss}) by first summing over the electron final states $\sum_{{\bf k}_f}\to({L}/{2\pi})^3\int d{\bf k}_f$ and subsequently dividing by the incoming plane wave particle flux $\hbar k_i / mL^3$. The total frequency-resolved scattering cross section is given by
\begin{equation}
    \sigma(\omega) = \frac{mL^3}{\hbar k_{i}}\Big(\frac{L}{2\pi}\Big)^3\int d\Omega_f \, k_{f}^2\,dk_{f}  \, w_{fi}(\omega),
\end{equation}
and by integration over frequency, the angle-resolved scattering cross section is
\begin{equation}
\begin{split}
    \frac{\partial \sigma}{\partial \Omega_f} &= \frac{mL^3}{\hbar k_{i}}\Big(\frac{L}{2\pi}\Big)^3\int d\omega \, k_{f}^2\, dk_{f} \, w_{fi}(\omega)\\
    &=-\frac{mL^3}{\hbar k_{i}}\Big(\frac{L}{2\pi}\Big)^3\int d\omega \, dE_{if} \, \bigg( \frac{\gamma_{f}m}{\hbar^2 k_f} \bigg) k_{f}^2 \, w_{fi}(\omega),
    \label{SCS_gen}
    \end{split}
\end{equation}
where $dE_{if} = -(\hbar^2 / m) k_f dk_{f}$ and $d\Omega_f=\sin\theta_f d\theta_f d\phi_f$. Lastly, noting that ${\partial\sigma}/ \partial \Omega_f=\int dE_{if}{\partial^2 \sigma}/({\partial E_{if} \partial \Omega_f})$, the double differential scattering cross section (DDCS) is defined as
\begin{widetext}
\begin{equation}
    \begin{aligned}
        \frac{\partial^2 \sigma}{\partial E_{if} \partial \Omega_f} & = -\frac{m^2 L^3}{\hbar^3} \Big(\frac{L}{2\pi}\Big)^3\bigg( \frac{ k_{f}}{k_i} \bigg) \int d\omega \, w_{fi}(\omega) \\
        & = \Big(\frac{m L^3}{\pi\hbar^2}\Big)^2 \bigg( \frac{ k_{f}}{k_i} \bigg) \int d\omega \, d\mathbf{x} d\mathbf{x}' \, \textrm{Im}\Big[ \mathbf{J}_{fi}^{*}(\mathbf{x}) \cdot \tensor{\mathbf{G}}(\mathbf{x}, \mathbf{x}', \omega) \cdot \mathbf{J}_{fi}(\mathbf{x}') \Big]\delta(\omega-\varepsilon_{if}),
    \end{aligned}
    \label{DDCS_final_form}
\end{equation}
\end{widetext}
which reduces to the DDCS common for an isolated dipolar target Eq. \eqref{DDCS_quasi} in core-loss EEL scattering \cite{Hebert2003-mn, PhysRevB.72.045142, bourgeois2023optical} when the electrostatic limit where $c\to\infty$ is taken. Further analytic progress is possible in the case of a single target dipole located at position $\mathbf{x}_d$ and characterized by frequency-dependent polarizability tensor $\tensor{\boldsymbol{\alpha}}(\omega)$. In this case, the induced Green's function reduces to $\tensor{\mathbf{G}}(\mathbf{x}, \mathbf{x}', \omega) = -({ 4 \pi\omega^2})\tensor{\mathbf{G}}^0(\mathbf{x}, \mathbf{x}_d, \omega) \cdot  \tensor{\boldsymbol{\alpha}}(\omega) \cdot \tensor{\mathbf{G}}^0(\mathbf{x}_d, \mathbf{x}', \omega)$, where $\tensor{\bf G}^0({\bf x},{\bf x}',\omega)=-1/({4 \pi \omega^2})[(\omega/c)^2\tensor{\bf I}+\nabla\nabla]{e^{i ({\omega}/{c}) |\mathbf{x} - \mathbf{x'}|}}/{|\mathbf{x} - \mathbf{x}'|}$ is the vacuum dipole Green's function, and the DDCS can be expressed analytically. Specifically, with a single incoming plane wave scattering to a single outgoing plane wave as described by the transition current density given by Eq. (\ref{J_fi_pw}), the fully retarded DDCS becomes
\begin{widetext}
\begin{equation}
    \frac{\partial^2 \sigma}{\partial E_{if} \partial \Omega_f} = \frac{ e^2}{\hbar^2 \pi}\bigg( \frac{ k_{f}}{k_i} \bigg)\frac{1}{\varepsilon_{if}^2} \, \textrm{Im}\bigg\{ \mathbf{Q} \cdot \bigg[ \bigg( \frac{\varepsilon_{if}}{c} \bigg)^2 \tensor{\mathbf{I}} - \mathbf{q} \mathbf{q} \bigg] \cdot  \frac{\tensor{\boldsymbol{\alpha}}(\varepsilon_{if})}{\big|({\varepsilon_{if}}/{c})^2 - q^2\big|^2} \cdot \bigg[ \bigg( \frac{\varepsilon_{if}}{c} \bigg)^2 \tensor{\mathbf{I}} - \mathbf{q} \mathbf{q} \bigg] \cdot \mathbf{Q} \bigg\},
    \label{DDCS_fullyret}
\end{equation}
\end{widetext}
where $\mathbf{Q} = 2\mathbf{k}_i - \mathbf{q}$. In the quasistatic limit ($c\to\infty$), Eq. (\ref{DDCS_fullyret}), reduces to the more familiar form \cite{SakuraiModern}
\begin{equation}
    \frac{\partial^2 \sigma}{\partial E_{if} \partial \Omega} =  \frac{ e^2}{\hbar^2 \pi}\bigg( \frac{ k_{f}}{k_i} \bigg)\frac{1}{\varepsilon_{if}^2} \, \textrm{Im}\bigg[\mathbf{Q} \cdot \mathbf{q} \mathbf{q} \cdot  \frac{\tensor{\boldsymbol{\alpha}}(\varepsilon_{if})}{q^4} \cdot \mathbf{q} \mathbf{q} \cdot \mathbf{Q} \bigg].
    \label{DDCS_quasi}
\end{equation}

\section{Numerical Implementation}
\label{sec_numerical}

This section details the numerical implementation of the inelastic scattering observables presented above in Secs. \ref{narrow_beam_section} and \ref{DDCS_section} for transversely phase-structured free electrons in both the focused beam and wide field limits. Our implementation generalizes the electron-driven discrete dipole approximation ($e$-DDA) \cite{Bigelow2012-pq, bigelow2013signatures}, built on top of the DDCSAT  \cite{draine2008discrete} framework, which previously utilized the vacuum electric field ${\bf E}^0_{fi}$ of a point electron as the source instead of an optical plane wave field. Other fully retarded and quasistatic numerical methods for simulating extended nanophotonic targets have been formulated to model low-loss electron beam interactions, such as the finite-difference time-domain \cite{talebi2013numerical,cao2015electron}, metal nanoparticle boundry element (MNPBEM) \cite{hohenester2012mnpbem}, and finite element \cite{doi:10.1021/nl3001309,pomplun2007adaptive} methods. In addition, transversely structured electron beams have been implemented in MNPBEM \cite{ugarte2016controlling,zanfrognini2019orbital,lourencco2021optical,Aguilar2023-zp}, albeit in the quasistatic limit only. Our treatment of the inelastic scattering of transversely structured electron beams in $e$-DDA is distinguished by its incorporation of fully retarded electron-sample interactions in both focused beam and wide field limits.

$e$-DDA/DDA originate from the method of coupled dipoles \cite{Purcell1973}, whereby the target is discretized into a finite collection of point electric dipoles 
\begin{equation}
{\bf p}_k(\omega)=\sum_{l=1}^N\Big[\tensor{\bm\alpha}(\omega)^{-1}-(-4\pi \omega^2)\tensor{\mathbf{G}}^0(\omega)\Big]^{-1}_{kl}\cdot\mathbf{E}_{fi}^{0}({\bf x}_l,\omega)
\label{ind_dipoles}
\end{equation}
of polarizability $\tensor{\bm\alpha}(\omega)$, each driven by the vacuum transition field $\mathbf{E}_{fi}^{0}({\bf x}_i,\omega)$ at frequency $\omega$ and mutually interacting via their fully-retarded electric dipole fields $-4 \pi \omega^2 \sum_l\tensor{\mathbf{G}}^0_{kl}(\omega)\cdot{\bf p}_l(\omega)$ until reaching self-consistency at that same frequency. Here $\tensor{\bf G}^0_{kl}(\omega)$ are the $kl$-matrix elements of the vacuum dipole Green's function $\tensor{\bf G}^0({\bf x},{\bf x}',\omega).$

Upon inversion of Eq. (\ref{ind_dipoles}), all EEL observables described above may be calculated from the resulting ${\bf p}_k(\omega)$ together with the applied field ${\bf E}^0_{fi}({\bf x}_k,\omega)$ (Eq. \eqref{vacE}) evaluated at each dipole. Specifically, the focused beam EEL probability and the wide field DDCS expressions in Eq. (\ref{EEL_probability}) and Eq. (\ref{DDCS_final_form}), respectively, can be adapted to a form that is compatible with the $e$-DDA code via the EEL rate per unit frequency 
\begin{widetext}
\begin{equation}
\begin{aligned}
    w_{fi}^{\textrm{loss}}(\omega)  &= -\frac{8 \pi}{\hbar} \int d\mathbf{x} d\mathbf{x}' \, \textrm{Im}\Big[ \mathbf{J}_{fi}^{*}(\mathbf{x}) \cdot \tensor{\mathbf{G}}(\mathbf{x}, \mathbf{x}', \omega) \cdot \mathbf{J}_{fi}(\mathbf{x}') \Big]\delta(\omega - \varepsilon_{if})\\
    &=\frac{2}{\hbar} \Big(\frac{v_i}{L}\Big)^2\,\textrm{Im} \Big[\sum_{k} \mathbf{E}^{0*}_{fi}(\mathbf{x}_k,\omega) \cdot  \mathbf{p}_k(\omega) \Big]  \delta(\omega - \varepsilon_{if}).
    \end{aligned}
    \label{wfi_CDA}
\end{equation}
\end{widetext}
Here the target's induced Green's function $\tensor{\bf G}(\mathbf{x},\mathbf{x}', \omega)$ is expanded in terms of the polarizabilities of the $N$ dipoles representing the target. That is 
\begin{widetext}
\begin{equation}
    \tensor{\bf G}(\mathbf{x},\mathbf{x}', \omega) = -{4 \pi\omega^{2} }\sum_{jj'}  \tensor{\bf G}^0(\mathbf{x}, \mathbf{x}_j, \omega) \cdot  \big[\tensor{\boldsymbol{\alpha}}^{-1}(\omega) -(- 4 \pi\omega^2 )\tensor{\bf G}^{0}(\omega) \big]^{-1}_{jj'} \cdot \tensor{\bf G}^0(\mathbf{x}_{j'}, \mathbf{x}', \omega),
\end{equation}
\end{widetext}
which can be derived from the response field of a polarizable body described in terms of its induced Green's function $\tensor{\bf G}(\mathbf{x},\mathbf{x}', \omega)$ driven by the external current density $\mathbf{J}^0(\mathbf{x})$ or by the free propagation (via $\tensor{\bf G}^0(\mathbf{x},\mathbf{x}', \omega)$) of the target's induced current density $\mathbf{J}(\mathbf{x})=\sum_j(-i\omega){\bf p}_j(\omega)\delta({\bf x}-{\bf x}_j)$ as ${\bf E}({\bf x},\omega)=-4\pi i \omega\int d{\bf x}' \, \tensor{\bf G}(\mathbf{x},\mathbf{x}', \omega)\cdot\mathbf{J}^{0}(\mathbf{x}')=- 4\pi i \omega\int d{\bf x}' \, \tensor{\bf G}^{0}(\mathbf{x},\mathbf{x}', \omega)\cdot\mathbf{J}(\mathbf{x}')$.

\subsection{Numerical evaluation of state- and energy-resolved EEL and EEG probabilities}
\label{numerical_narrow_beam_section}
From Eq. (\ref{wfi_CDA}), the state- and energy-resolved EEL probability can be obtained following the same procedure in Sec. \ref{narrow_beam_section}. Specifically, the transversely phase-shaped EEL probability becomes
\begin{equation}
    \Gamma_{fi}^{\textrm{loss}}(\omega) =\frac{1}{\pi\hbar^2} \textrm{Im} \Big[\sum_{k} \mathbf{E}^{0*}_{fi}(\mathbf{x}_k,\omega =  q_{\parallel}v_i) \cdot  \mathbf{p}_k(\omega = q_{\parallel}v_i) \Big], 
   \label{Gamma_fi_DDA}
\end{equation}
when working within the nonrecoil approximation introduced in Sec. \ref{sec_observables}, as appropriate to focused beams prepared in the STEM configuration. The vacuum transition electric field $\mathbf{E}^{0}_{fi}$ appearing within Eq. \eqref{ind_dipoles} and Eq. \eqref{Gamma_fi_DDA} can, in principle, be any of the fields sourced by the transversely focused transition current densities described in Sec. \ref{sec_Jfi}. However, due to their complexity, only those transitions involving OAM transfers $\Delta\ell=1$ depicted as colored arrows in Fig. \ref{F2}(e), and more specifically the resulting electric fields sourced by the transition current densities seen in Fig. \ref{F3}(c) and \ref{F3}(d), have been implemented within the $e$-DDA code. Explicit forms for these fields are provided in the Appendix. Phase-shaped EEG probability spectra of nanophotonic targets under continuous-wave laser stimulation can also be evaluated numerically using the $e$-DDA as detailed previously \cite{bourgeois2022polarization}. Briefly, the EEG probability in Eq. (\ref{EEG_NBWL_final_form}) is numerically integrated by quadrature using the optically-induced response field of the target $\mathbf{E}_{\nu}^{(+)}$ calculated using DDCSAT and the $\hat{\mathbf{J}}_{fi}$ defined by selection of a specific pair of incoming and outgoing free electron states. 

\begin{figure}
\includegraphics[]{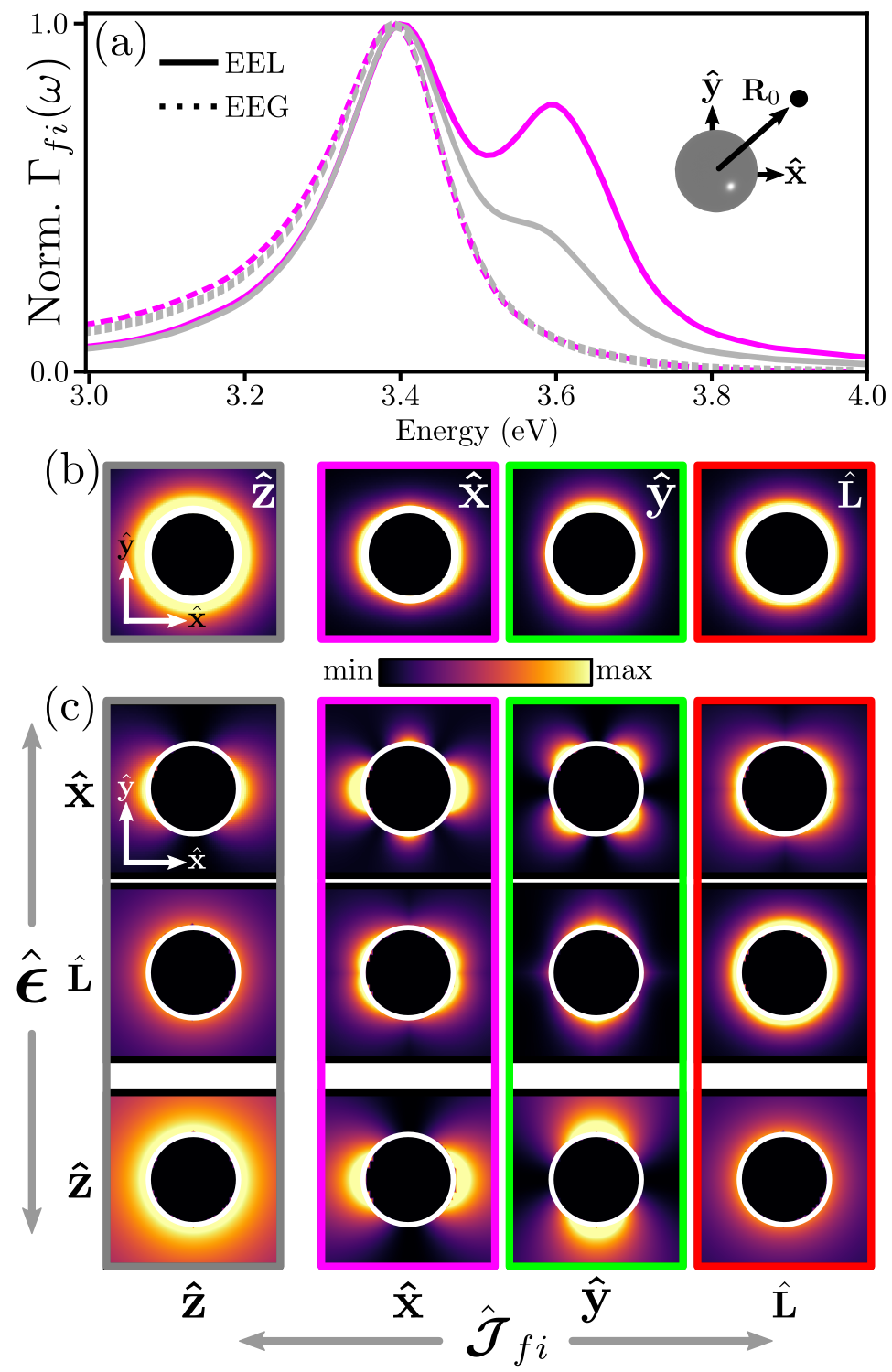}
    \caption{Phase-shaped EEL and EEG point spectra and spectrum images of a silver plasmonic sphere of 25 nm radius evaluated using $e$-DDA. (a) Phase-shaped EEL (solid) and EEG (dashed) point spectra, evaluated in the narrow beam limit, at impact parameter $\mathbf{R}_0 = (1/\sqrt{2}) (40\textrm{ nm},40\textrm{ nm})$. The optical polarization is $\hat{\boldsymbol{\epsilon}} = \hat{\mathbf{x}}$ in the laser-stimulated EEG calculations. Gray and magenta colors correspond to $\hat{\boldsymbol{\mathcal{J}}}_{fi}$ along $\hat{\mathbf{z}}$ and $\hat{\mathbf{x}}$, respectively. (b) Phase-shaped EEL spectrum images at the dipole LSP resonance energy of 3.4 eV. Each column corresponds to the labeled  $\hat{\boldsymbol{\mathcal{J}}}_{fi}$. (c) Phase-shaped EEG spectrum images at 3.4 eV. The optical excitation polarization state $\hat{\boldsymbol{\epsilon}}$ varies between rows, while each column corresponds to the labeled $\hat{\boldsymbol{\mathcal{J}}}_{fi}$. For $\hat{\boldsymbol{\epsilon}} = \{ \hat{\bf{x}}, \hat{\bf{L}}\}$, the optical axis is chosen to be along $\hat{\bf{z}}$, while for $\hat{\boldsymbol{\epsilon}} = \bf{\hat{z}}$ it is chosen to be along $\hat{\bf{x}}$. Panels with $\hat{\boldsymbol{\mathcal{J}}}_{fi} = \{ \hat{\mathbf{x}},\,\hat{\mathbf{y}},\hat{\mathbf{L}} \}$ and with $\hat{\boldsymbol{\epsilon}} = \{\hat{\mathbf{x}},\hat{\mathbf{L}} \}$ share a common scale, while panels with $\hat{\boldsymbol{\mathcal{J}}}_{fi} =\bf{\hat{z}}$ are separately scaled. The incident velocity of the probing electron is ${\bf v}_i=0.7c\,\hat{\bf z}$ in all cases.
    }
    \label{F4}
\end{figure}

Figure \ref{F4} presents a comparison of normalized focused beam EEL and EEG spectra for a 25 nm radius silver sphere calculated using $e$-DDA with dielectric data taken from Ref. \cite{PhysRevB.6.4370}. Probing electrons have a speed of 0.7$c$.  Panel (a) shows phase-structured EEL (solid traces) and EEG (dashed traces) spectra at impact parameter $\mathbf{R}_0 = (1/\sqrt{2}) (40\textrm{ nm},40\textrm{ nm})$. The optical excitation polarization $\hat{\boldsymbol{\epsilon}}$ in the EEG calculations is along $\hat{\mathbf{x}}$. Gray and magenta colors correspond to pre- and post-selection of HG transitions $\Psi_{00}(x,y) \rightarrow \Psi_{00}(x,y)$ with $\hat{\boldsymbol{\mathcal{J}}}_{fi}^\parallel = \hat{\mathbf{z}}$, and $\Psi_{10}(x,y)\rightarrow\Psi_{00}(x,y)$ with $\hat{\boldsymbol{\mathcal{J}}}^{\perp}_{fi} = \hat{\mathbf{x}}$, respectively. While the $e$-DDA calculations can capture coupling to the higher-order multipoles, as seen in Fig. \ref{F4}(a), the dipolar localized surface plasmon (LSP) is clearly evident near 3.4 eV. Here it is apparent that both transitions with $\hat{\boldsymbol{\mathcal{J}}}_{fi}$ oriented parallel and perpendicular to the electron trajectory couple to the quadrupolar LSP mode near 3.6 eV, albeit with different strengths. The EEG spectra, meanwhile, are dominated by the optically bright dipolar response of the sphere, whereas the higher-order, dark modes are inaccessible by the stimulating optical pump field. 

EEL spectrum images obtained by plotting $\Gamma_{fi}^{\textrm{loss}}(\mathbf{R}_0, \omega)$ as a function of the impact parameter ${\bf R}_0$ at the dipole LSP energy 3.4 eV are shown in Fig. \ref{F4}(b). In the conventional EEL case, where $\hat{\boldsymbol{\mathcal{J}}}_{fi}^{\perp} = {\bf 0}$ resulting in $\hat{\boldsymbol{\mathcal{J}}}_{fi} =\hat{\boldsymbol{\mathcal{J}}}_{fi}^{\parallel} = \hat{\mathbf{z}}$ (gray, left), the spectrum image exhibits the expected circular symmetry. Meanwhile, when $\hat{\boldsymbol{\mathcal{J}}}_{fi}^{\perp} \neq {\bf 0}$, producing a $\hat{\boldsymbol{\mathcal{J}}}_{fi}^{\perp}$ oriented along $\hat{\mathbf{x}}$ ($\hat{\mathbf{y}}$), and outlined in magenta (green), the spectrum image shows slight elongation along the direction of the transition current density. As required by symmetry, the circularly-polarized $\hat{\boldsymbol{\mathcal{J}}}_{fi}^{\perp} = \hat{\mathbf{L}}$ (red) couples with radial symmetry to the spherical target. In Fig. \ref{F4}(b), the $\hat{\boldsymbol{\mathcal{J}}}^\parallel_{fi} = \hat{\mathbf{z}}$ plot is normalized to a maximum of $10^{-2}$ eV$^{-1}$, while the $\hat{\boldsymbol{\mathcal{J}}}_{fi}^{\perp} \in \{ \hat{\mathbf{x}}, \hat{\mathbf{y}}, \hat{\mathbf{L}} \}$ cases share a common normalization factor of $10^{-8}$ eV$^{-1}$ for a beam waist of 1 nm. The small perpendicular to parallel ratio of signals represents a hurdle to phase-shaped EEL spectroscopy measurements regarding limits of detection, although measurements of this type have been achieved previously \cite{guzzinati2017probing}. Phase-shaped EEG spectrum images at the same dipole LSP energy 3.4 eV are presented in Fig. \ref{F4}(c). The optical excitation polarization state $\hat{\boldsymbol{\epsilon}}$ varies between rows, while each column corresponds to the labeled component of $\hat{\boldsymbol{\mathcal{J}}}_{fi}$. For $\hat{\boldsymbol{\epsilon}} = \{ \hat{\bf{x}}, \hat{\bf{L}}\}$, the optical axis is chosen to be along $\hat{\bf{z}}$. There $\hat{\boldsymbol{\mathcal{J}}}_{fi}^{\perp} =\mathbf{\hat{x},\,\hat{y},\,\textrm{and }\hat{L}}$ share a common scale factor, while $\hat{\boldsymbol{\mathcal{J}}}_{fi} =\bf{\hat{z}}$ is separately scaled. The ratio of the transverse to longitudinal EEG probabilities is $\Gamma_{{fi}\perp}^{\textrm{gain}} / \Gamma_{{fi}\parallel}^{\textrm{gain}} \approx 10^{-5}$, for a 200 keV electron beam with a waist 1 nm. These findings are consistent with earlier theoretical \cite{bourgeois2022polarization} and experimental \cite{li2019plasmon} studies. For the $\hat{\boldsymbol{\epsilon}} = \bf{\hat{z}}$ case, the optical axis is chosen to be along $\hat{\bf{x}}$ and the ratio of transverse to longitudinal EEG signals remains similarly small ($\sim 10^{-5}$), but all signals are smaller by $\sim 10^{-3}$ in this excitation geometry. Comparison of Figs. \ref{F4}(b) and \ref{F4}(c) highlights the differing identities of the excitation sources and roles played by $\mathbf{J}_{fi}$ in EEL and laser-stimulated EEG processes. When considering EEL events, the STEM electron acts as both a spatially dependent pump and probe sourced by the transition current density $\mathbf{J}_{fi}(\mathbf{R}_0)$ at impact parameter $\mathbf{R}_0$. In stark contrast, as alluded to in Sec. \ref{theory_IES}, the pump and the probe are decoupled for laser-stimulated EEG processes. Specifically, as seen in Eq. (\ref{w_fi_gain}), the pump is the optical source exciting the target's induced field ${\bf E}^{(+)}_{\nu}$, which is then probed by the electron's transition current density $\mathbf{J}_{fi}(\mathbf{R}_0)$ at $\mathbf{R}_0$.

\subsection{Numerical evaluation of the inelastic double differential cross section}
\label{numerical_wide_beam_section}
From Eqs. \eqref{DDCS_final_form} and \eqref{wfi_CDA}, the wide field inelastic DDCS can be expressed as
\begin{widetext}
\begin{equation}
    \begin{aligned}
        \frac{\partial^2 \sigma}{\partial E_{if} \partial \Omega_f} & = -\frac{ m^2 L^6}{(2\pi)^3\hbar^3} \bigg( \frac{ k_{f}}{k_i} \bigg) \int d\omega \, w_{fi}^{\textrm{loss}}(\omega) \\
        & = - \frac{1}{4 \pi^3} \frac{ k_{f}}{k_i} \bigg(\frac{v_i m L^2}{\hbar^2}\bigg)^2 \textrm{Im} \Big[\sum_{j} \mathbf{E}^{0*}_{fi}(\mathbf{x}_j,\omega = \varepsilon_{if}) \cdot  \mathbf{p}_j(\omega = \varepsilon_{if}) \Big].
    \end{aligned}
\end{equation}
\end{widetext}
We have implemented within $e$-DDA the plane wave transition fields defined in Appendix Eqs. (\ref{Efi_pw_single}) and (\ref{Efi_pw_superposition}), sourced by the transition currents Eqs. (\ref{J_fi_pw}) and (\ref{J_fi_pw_super}), respectively, for the cases of scattering from either a single or a coherent superposition of two incident plane waves, to a single outgoing plane wave state.

\begin{figure*}
\includegraphics{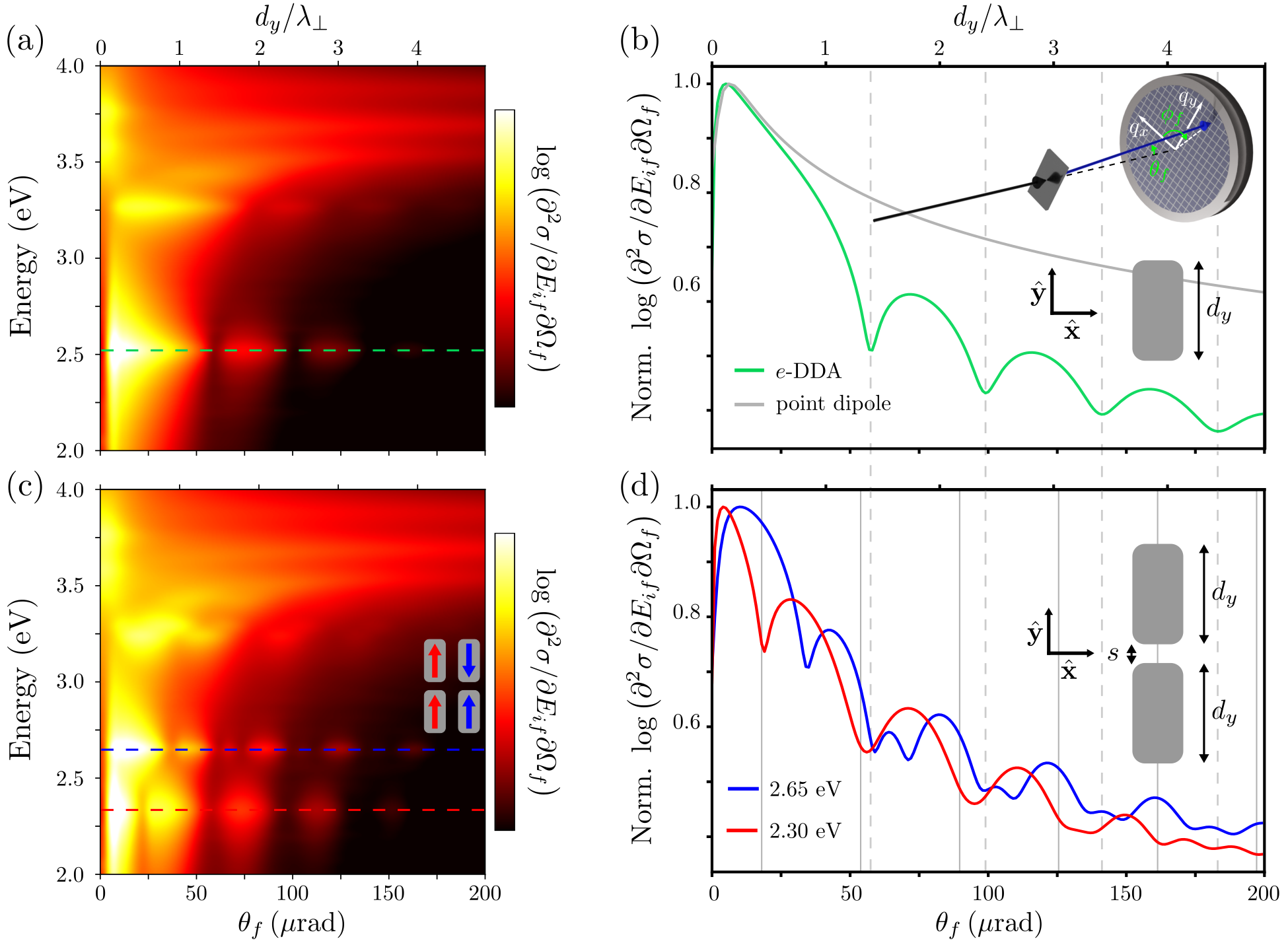}    
    \caption{DDCSs for plasmonic nanorod monomer and dimer systems evaluated using $e$-DDA. (a) Numerically calculated DDCS (logarithmic scale) for a silver nanorod with Cartesian dimensions of 30 nm $\times$ 60 nm $\times$ 15 nm. The incident electron is directed along $\hat{\mathbf{z}}$ with 200 keV kinetic energy and the collected scattering direction is varied from $\theta_f=0$ $\mu$rad (along $\hat{\bf z}$) to $\theta_f = 200$ $\mu$rad, with $\phi_f=\pi/2$ fixed such that $\mathbf{q}_{\perp} = q_\perp \hat{\mathbf{y}}$. The quantity $d/\lambda_{\perp}$ (upper horizontal axis) is the rod length ($60$ nm) along the momentum recoil direction divided by the transverse recoil wavelength $\lambda_\perp= 2 \pi/q_\perp$. (b) Lineout of the DDCS in panel (a) at 2.5 eV (green). The gray trace is the DDCS of an anisotropic point dipole evaluated using the analytic expression in Eq. \eqref{DDCS_fullyret}. Vertical gray dashed lines indicate single particle diffraction minima. Insets show schematic depictions of the scattering and target geometries. (c) Same as (a), but for a homodimer target composed of two copies of the nanostructure considered in (a) and (b)  arranged tip-to-tip along the $y$-axis with an $s = 10$ nm gap. (d) Lineouts from (c) at 2.30 eV (red) and 2.65 eV (blue). Vertical gray dashed lines indicate single slit diffraction minima, while solid gray lines denote double slit interference minima at 2.30 eV.}
    \label{F5}
\end{figure*}

The DDCSs for inelastic scattering of wide field plane wave electron states from plasmonic nanorod monomers and dimers are presented in Fig. \ref{F5}. Probing electrons have an initial kinetic energy of 200 keV and wave vector directed along the TEM axis, i.e., $\hat{\mathbf{k}}_i = \hat{\mathbf{z}}$, while the outgoing wave vectors $\mathbf{k}_f$ possess a nonzero $y$-component such that the transverse momentum recoil $\hbar \mathbf{q}_{\perp}$ is along $\hat{\mathbf{y}}$. Working in the low loss energy regime in order to observe the plasmonic modes of interest, Fig. \ref{F5}(a) shows the calculated DDCS (logarithmic scale) for an anisotropic silver rod with Cartesian dimensions 30 nm $\times$ 60 nm $\times$ 15 nm (width, length, height) as a function of scattering angle $\theta_f$ and loss energy. The nanorod is orientated with its longest dimension $d_y$ along the direction of collected transverse recoils ($\phi_f=\pi/2$), leading to the lowest energy long-axis dipole mode near 2.50 eV dominating the DDCS. Since the transversely polarized components of the transition field $\mathbf{E}^0_{fi}$ in Eq. \eqref{Efi_pw_single} are proportional to $\mathbf{q}_{\perp}$, transversely oriented LSP modes of the nanorod do not contribute to the DDCS at $\theta_f=0$ $\mu$rad. The apparent features at $\theta_f=0$ $\mu$rad and energies above 3.50 eV in Fig. \ref{F5}(a) arise from longitudinal multipoles oriented along $\hat{\mathbf{z}}$.

Figure \ref{F5}(b) shows a lineout from panel (a) at the long-axis dipole LSP energy marked by the dashed green line. The gray trace in Fig. \ref{F5}(b) is calculated using the analytic form of the DDCS given in Eq. \eqref{DDCS_fullyret} for a single point dipole representing the target. The anisotropy of the nanorod response is captured by detuning the shorter axis dipole LSP energies above 2.50 eV in the dipole's effective polarizability. As a consequence,  $\partial^2 \sigma/\partial E_{if} \partial \Omega_f \propto |\mathbf{E}^0_{fi} \cdot \hat{\mathbf{y}}|^2$, which ensures $\partial^2 \sigma/\partial E_{if} \partial \Omega_f |_{\theta_f=0} =0$. At moderate opening angles ($\gtrsim 5$ $\mu$rad) the DDCS decreases as the opening angle increases, which starting from Eq. \eqref{Efi_pw_single} and  $Q_\parallel \gg |{\bf Q_\perp}| = |{\bf q_\perp}|$ can be shown by  ${\bf E}^0_{fi}  \cdot \hat{\mathbf{y}}\propto Q_\parallel q_\parallel {\bf q}_\perp \cdot \hat{{\bf y}} / \big({\bf q}_{\perp}^2 + q_{\parallel}^2 - \omega^2/c^2\big)  = Q_\parallel q_\parallel |{\bf k}_f| \sin \theta_f/ \big[|{\bf k}_f|^2\sin^2 \theta_f + \omega^2(1/v_i^2 - 1/c^2) \big]$. For a 200 keV  electron with loss energy $\hbar\epsilon_{if} = 2.5$ eV,  $|{\bf k}_f| \sin \theta_f \gtrsim \omega/(\gamma_i v_i)$ at opening angles $> 5\,\mu\textrm{rad}$, leading to $\partial^2 \sigma/\partial E_{if}d\Omega_f\,{\propto}\,1/\sin^2\theta_f$.  This effect exists independent of the target geometry and represents the decreasing probability of low loss events with moderate transverse recoil. Separately the lineouts in $\theta_f$ on the order of 5 $\mu$rad range are primarily dictated by the growing in of the transverse LSP mode, as at smaller angles $\mathbf{E}^0_{fi} \cdot \hat{\mathbf{y}}\propto {\bf q} \cdot \hat{\mathbf{y}} =|{\bf k}_f|\sin \theta_f$ and the lineout is taken at the ${\bf\hat{y}}$ oriented dipole mode energy. Unlike in the narrow beam limit, the DDCS observable has equal magnitude contributions from transverse and longitudinal recoils, which is a well known experimental and theoretical result of EEL DDCS on an anisotropic target \cite{hebert2006elnes,PhysRevB.72.045142}. In addition to tracking the angular scattering behavior predicted by the point dipole model at small scattering angles ($ \lesssim25$ $\mu$rad), the lineout shown in green exhibits a progression of diffraction maxima/minima with increasing $\theta_f$ arising from the finite extent of the target. The single particle diffraction minima, indicated by vertical gray dashed lines, are located nearby angles $\theta_m$ corresponding to single slit diffraction minima predicted using $d_y \sin \theta_m = m \lambda_e$, where integers $m$ index the diffraction minima, and $\lambda_e$ is the de Broglie wavelength of the electron. 

Figures \ref{F5}(c) and \ref{F5}(d) consider a nanorod dimer consisting of a pair of the silver rods from Figs. \ref{F5}(a) and \ref{F5}(b) displaced along $\hat{\mathbf{y}}$ such that there is an $s=10$ nm gap between the rod tips. The dimer's DDCS is presented in Fig. \ref{F5}(c), where the bonding (red) and antibonding (blue) hybridized long-axis dipole LSP modes are visible at energies slightly below and above 2.5 eV, respectively. Lineouts from Fig. \ref{F5}(c) at loss energies corresponding to the bonding and antibonding dimer modes are shown in Fig. \ref{F5}(d), again displaying multiple diffraction minima/maxima. In analogy to the double slit experiment, each nanorod is a source of single-particle diffraction in addition to matter-wave interference arising from the $d_\perp + s = 70$ nm center-to-center displacement of the the nanorods. For example, each DDCS minimum observed at the bonding mode energy (2.30 eV) occurs at an angle corresponding to either one of the single-particle diffraction minima from Fig. \ref{F5}(b) marked by vertical gray dashed lines, or at angles $\theta_n$ satisfying the double slit interference condition $n\lambda_\perp=d_y + s$, which are indicated by vertical gray solid lines.  The condition for constructive interference at the antibonding resonance energy is $(n+1/2)\lambda_\perp=d_y + s$.

\begin{figure}
\includegraphics[]{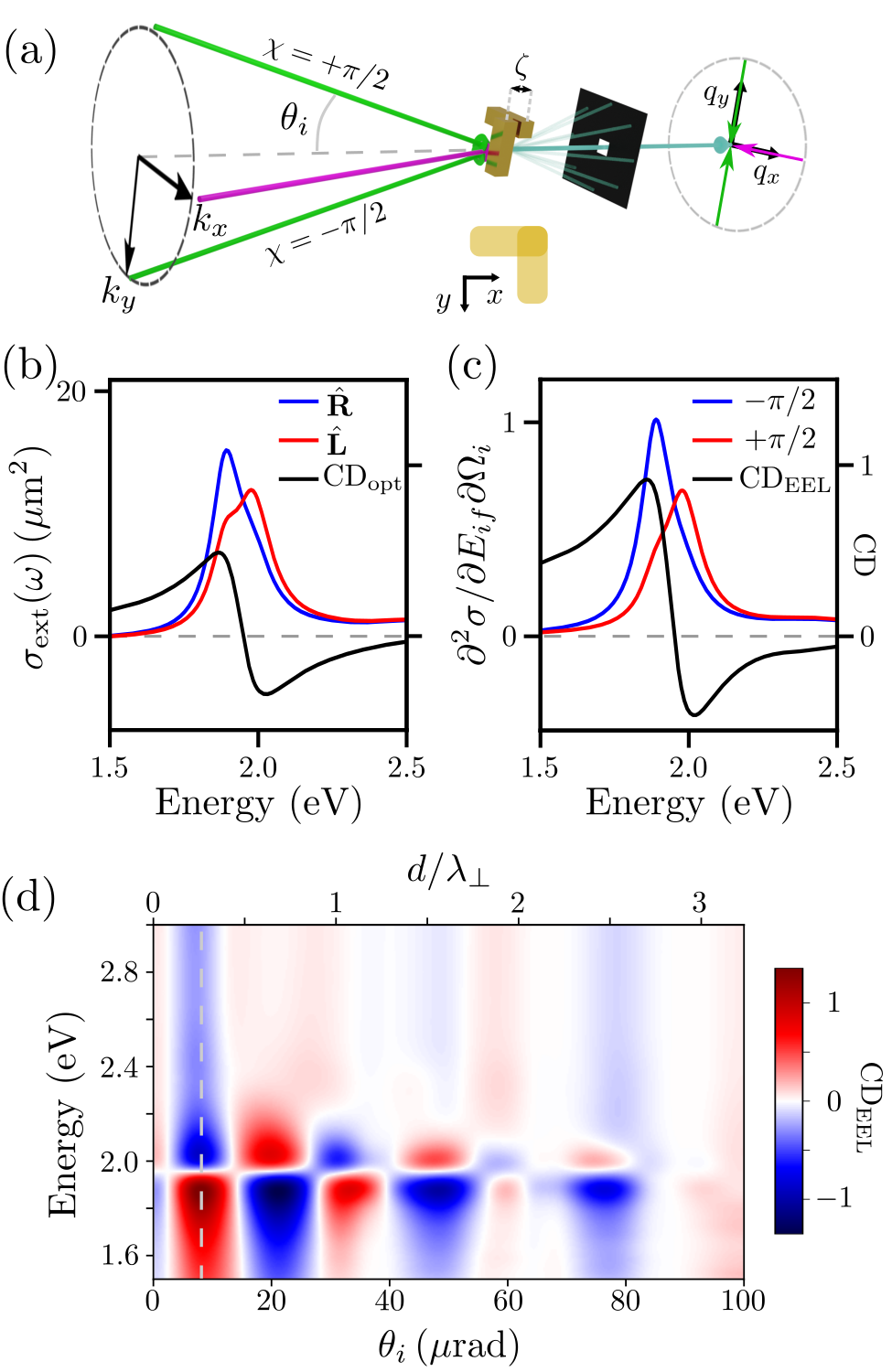} 
\caption{Numerical evaluation of the circular dichroic response of a chiral nanorod structure probed using wide field EEL spectroscopy within $e$-DDA. (a) Scheme showing the gold nanorod BK structure and probing geometries. Superposition plane wave electron states are incident from the left, while an aperture (black) post-selects the outgoing electron state along the TEM ($\hat{\mathbf{z}}$) axis. (b) Optical extinction spectra for incident plane wave wave vectors along $\hat{\mathbf{z}}$ with circular polarization states $\hat{\boldsymbol{\epsilon}} = \hat{\mathbf{R}}$ (blue) and $\hat{\mathbf{L}}$ (red). The optical circular dichroism CD$_{\textrm{opt}}$ is shown in black and shares its secondary $y$-axis with panel (c). (c) DDCS spectra of the same BK structure for superposition plane wave electron states with $\chi= + \pi/2$ (red) and $\chi= - \pi/2$ (blue). The CD$_{\textrm{EEL}}$ spectrum is shown in black. (d) Angle $\theta_i$-resolved CD$_{\textrm{EEL}}$ spectra. The dashed gray line in (d) indicates the fixed angle of the lineouts in (c). The incident electron kinetic energy in (c,d) is 200 keV.}    
\label{fig_bks}
\end{figure}

An investigation of inelastic scattering involving superposition plane wave electron states interacting with a chiral nanophotonic target is presented in Fig. \ref{fig_bks}. Specifically, the target is the well-understood Born-Kuhn structure (BK) \cite{yin2013interpreting}, composed of two gold nanorods arranged in an L-shape with a relative displacement $\zeta$ between the rods centered along $\hat{\mathbf{z}}$, as demonstrated in Fig. \ref{fig_bks}(a). Here, the long and short axes of each nanorod are 100 nm and 30 nm, respectively, $\zeta = 40$ nm, and tabulated gold dielectric data are taken from the literature \cite{PhysRevB.6.4370}. Fig. \ref{fig_bks}(b) shows the optical extinction spectra of the BK target for incident wave vectors along $\hat{\mathbf{z}}$ and polarizations $\hat{\boldsymbol{\epsilon}} = \hat{\mathbf{L}}$ (red) and $\hat{\mathbf{R}}$ (blue). As is well-understood (see Appendix) \cite{fu2017chiral, warning2021nanophotonic} both incident field helicities couple, albeit with unequal strengths, to the bonding and antibonding LSP modes near 2.0 eV involving the long-axis dipoles. The optical circular dichroism $\textrm{CD}_\textrm{opt} = 2(\sigma^{\hat{\mathbf{R}}}_{\textrm{ext}} - \sigma^{\hat{\mathbf{L}}}_{\textrm{ext}})/(\sigma^{\hat{\mathbf{R}}}_{\textrm{ext}} + \sigma^{\hat{\mathbf{L}}}_{\textrm{ext}})$ spectrum is shown in black, which exhibits the BK system's chiral response and serves as a point of comparison against the electron-based observables.

As depicted in Fig. \ref{fig_bks}(a), free electron plane wave superposition states $\psi_{k_xk_y}^{\chi}(x,y,z)$ are incident from the left, while an aperture situated in the diffraction plane post-selects the outgoing free electrons along the TEM axis with $\mathbf{k}^f_{\perp}={\bf 0}$, i.e, $\psi_{k'_\parallel}(x,y,z)$. It has been shown previously that the response of targets as measured by wide field EEL spectroscopy with pre- and post-selection of such states with $k_x = k_y$ and $\chi = \pm \pi/2$ in many ways mimic the optical response under circularly polarized optical excitation \cite{Hebert2003-mn, Schattschneider2006-fy, bourgeois2023optical}. Indeed, Fig. \ref{fig_bks}(c) presents DDCS spectra of the BK target for $\psi_{k_xk_y}^{\chi}(x,y,z) \to \psi_{k_\parallel'}(x,y,z)$, producing the transition current density given by Eq. (\ref{J_fi_pw_super}), for the scenarios of $\chi = \pm \pi/2 $. The circular dichroism CD$_{\textrm{EEL}}$ (black) is again defined as twice the difference in the $\chi = \pm \pi/2$ DDCS spectra normalized by the sum, and strongly resembles the CD$_{\textrm{opt}}$ spectrum in Fig. \ref{fig_bks}(b). Panel (c) presents the CD$_{\textrm{EEL}}$ spectra as a function of convergence angle $\theta_i$. Like the nanorod dimer presented in Fig. \ref{F5}, the BK system exhibits both single slit diffraction and double slit interference effects (Fig. \ref{fig_bks}(d)), though the superposition state excitation and BK target geometry conspire to produce more complicated beating patterns than those observed in Fig. \ref{F5}(c). The vertical dashed gray line in Fig. \ref{fig_bks}(d) indicates the convergence angle $\theta_i$ of the spectra presented in panel (c).

\section{Conclusions}
Development of capabilities to prepare, parse, and measure transversely phase-structured electron states in the electron microscope has opened the door to state- and energy-resolved inelastic scattering observables, adding to the rapidly evolving toolkit used to interrogate nanoscale systems in the low-loss regime. Here we present general expressions for transversely phase-shaped EEL and continuous-wave laser-stimulated EEG spectroscopies for transversely localized vortex and non-vortex electron states, and the DDCS for wide field electron plane waves under minimal assumptions regarding the magnitudes of electron velocity and energy exchange. By exploiting a quantum mechanical treatment that accounts explicitly for the transverse degrees of freedom of the probing electron wave functions, we have showcased the ability to retrieve information about the optical near-field and electromagnetic response of nanophotonic targets using inelastic scattering of phase-shaped free electrons following energy-momentum post-selection.  A numerical procedure for evaluating derived observables is presented that allows for flexibility regarding particle number, size, geometry, and material composition. Example calculations for several prototypical plasmonic monomer and dimer systems are investigated to highlight the utility of our approach to analyze mode symmetries, local response field characteristics, chiral responses, and matter-wave diffraction phenomena. The general procedures outlined for constructing wide field plane wave and nondiffracting twisted and non-twisted electron beams with distinct transverse polarization and topological textures, in addition to the state- and energy-resolved observables, can be readily applied to many areas of atomic, molecular, and materials physics. In particular, we have drawn attention to an application of free electron qubits, whereby transitions between different OAM qubit states produce transition current densities with unique polarization and vector profiles, including analogs of optical polarization states and other more general forms of structured light. The theoretical framework presented can be extended to describe beam coherence in electron holography \cite{R_Harvey2014-te, Yasin2018-kb} via a density matrix formalism \cite{SciPostPhys.10.2.031}, and utilized to explore the role of pure and mixed electron states in inelastic scattering and the concurrent transfer of quantum information in the form of quantized energy and OAM \cite{10.1126/sciadv.abe4270, PhysRevX.13.031001, Loffler2022-lh, L_ffler_2023}. Furthermore, the use of transversely phase-structured free electron states realizable in TEMs, STEMs, and ultrafast TEMs can lead to additional manifestations of unique electromagnetic fields \cite{PhysRevA.91.033808, PhysRevA.95.012106} and electron paraxial skyrmionic beams \cite{PhysRevA.102.053513}, all of which could play vital roles in the investigation of novel optically forbidden atomic, molecular, material, and topological excitations \cite{PhysRevA.108.043513, Davis2020-kr, Shen2022-uz, PhysRevLett.131.196801, Ghosh2023-ab}.

\begin{acknowledgments}
All work was supported by the U.S. Department of Energy (DOE), Office of Science, Office of Basic Energy Sciences (BES), Materials Sciences and Engineering Division under Award No. DOE BES DE-SC0022921. 
\end{acknowledgments}

\section{Appendix}
\subsection{Vacuum Transition Vector Potentials and Electric Fields}
The transition vector potentials and electric fields associated with each of the transition current densities in Sec. \ref{sec_Jfi} are presented below. Specifically, the vacuum electric field 
\begin{equation}
\begin{split}
    {\bf E}^0_{fi}({\bf x},\omega) &= \frac{i c}{\omega}\bigg[\Big(\frac{\omega}{c}\Big)^2{\tensor{\bf I}}+{\nabla\nabla}\bigg]\cdot{\bf A}^0_{fi}({\bf x},\omega) \\
    &= \frac{i}{\omega} \int d\mathbf{x}' \,  \bigg[\Big(\frac{\omega}{c}\Big)^2{\tensor{\bf I}}+\nabla\nabla\bigg]\frac{e^{i ({\omega}/{c}) |\mathbf{x} - \mathbf{x'}|}}{|\mathbf{x} - \mathbf{x}'|}\,\Big(\frac{L}{v_i}\Big)\mathbf{J}_{fi}(\mathbf{x}')\\
    &=-4\pi i\omega \int d\mathbf{x}' \, \tensor{\mathbf{G}}^0(\mathbf{x}, \mathbf{x}', \omega)\cdot\Big(\frac{L}{v_i}\Big)\mathbf{J}_{fi}(\mathbf{x}')
    \end{split}
    \label{vacE}
\end{equation}  
plays an important role in the observable EEL and EEG processes and depends upon the vector potential
\begin{equation}
    \mathbf{A}^0_{fi}(\mathbf{x},\omega) =  \frac{1}{c}\int d\mathbf{x}' \,  \frac{e^{i ({\omega}/{c}) |\mathbf{x} - \mathbf{x'}|}}{|\mathbf{x} - \mathbf{x}'|}\,\Big(\frac{L}{v_i}\Big)\mathbf{J}_{fi}(\mathbf{x}')
    \label{induced_vecpot}
\end{equation}
presented here in the Lorenz gauge. Beginning with the transition current density in Eq. (\ref{J_fi_pw}) associated with the transition between single plane wave states $\psi_{{\bf k}_i}(x,y,z)$ and $\psi_{{\bf k}_f}(x,y,z)$ where ${{\bf k}_i}$ and ${{\bf k}_f}$ are arbitrary wave vectors, the vacuum vector potential 
\begin{equation}
    \mathbf{A}^0_{{{\bf k}_f}{{\bf k}_i}}(\mathbf{x},\omega) = \frac{2 \pi e\hbar}{m c v_i L^2} \frac{e^{i \mathbf{q}\cdot \mathbf{x}}}{\big({\omega}/{c}\big)^2 - q^2}{\bf Q}
\end{equation}
and electric field 
\begin{equation}
    \mathbf{E}^0_{{{\bf k}_f}{{\bf k}_i}}(\mathbf{x},\omega) = \frac{2\pi ie \gamma_i}{k_i L^2 \omega}  \frac{e^{i \mathbf{q}\cdot \mathbf{x}}}{({\omega}/{c})^2 - q^2} \Big[\Big(\frac{\omega}{c}\Big)^2\tensor{\bf{I}} - \mathbf{q}\mathbf{q} \Big] \cdot \mathbf{Q}
    \label{Efi_pw_single}
\end{equation}
are readily obtained upon using the integral identity $\int d\mathbf{x}' {e^{i(\omega / c)|\mathbf{x} - \mathbf{x}'|}}e^{\mp i\mathbf{q}\cdot \mathbf{x}'}/{|\mathbf{x} - \mathbf{x}}'|=-4 \pi e^{\mp i\mathbf{q}\cdot \mathbf{x}}/[(\omega/c)^2 - q^2]$.

The vacuum transition vector potential sourced by the superposition plane wave state transition current density in Eq. (\ref{J_fi_pw_super}) takes the form
\begin{widetext}
\begin{equation}
    \begin{aligned}
        \mathbf{A}_{k_\parallel',k_xk_y}^{ \chi 0}(\mathbf{x},\omega) & = \frac{2\pi e\hbar}{ \sqrt{2}m c v_i L^2} e^{iq_{\parallel}z}\bigg[ \frac{k_x e^{ik_x x}}{({\omega}/{c}\big)^2 - k_x^2 - q_{\parallel}^2}\hat{\mathbf{x}} + \frac{k_y e^{i\chi} e^{ik_y y}}{({\omega}/{c})^2 - k_y^2 - q_{\parallel}^2}\hat{\mathbf{y}} \\
        &\ \ \ + Q_{\parallel}\bigg(\frac{ e^{ik_x x}}{({\omega}/{c})^2 - k_x^2 - q_{\parallel}^2} + \frac{e^{i\chi} e^{ik_y y}}{({\omega}/{c})^2 - k_y^2 - q_{\parallel}^2} \bigg) \hat{\mathbf{z}} \bigg]
    \end{aligned}
\label{Afi_pw_superposition}
\end{equation}
\end{widetext}
for the superposition state $\psi_{k_xk_y}^{\chi}(x,y,z) = (1 / \sqrt{2L})\big[\Psi_{k_x}(x,y) + \Psi_{k_y}(x,y)e^{i\chi}\big]e^{ik_{\parallel}z}$  introduced in Sec. \ref{electron_states_section}, transitioning to the final pin hole state $\psi_{k_{\parallel}'}(x,y,z)=L^{-3/2}e^{ik'_{\parallel}z}$ oriented along the TEM ($\hat{\bf z}$) axis. Associated with this vector potential is the superposition plane wave state transition electric field 
\begin{widetext}
\begin{equation}
    \begin{aligned}
        {\bf\mathbf{E}}^{\chi 0}_{k_\parallel',k_xk_y}(\mathbf{x},\omega) = \frac{-2\pi i e \hbar}{\sqrt{2}m \omega L^2 v_i}\bigg[&\frac{k_x^2  + Q_\parallel q_\parallel - (\frac{\omega}{c})^2}{k_x^2 + q_\parallel^2 - (\frac{\omega}{c})^2}k_xe^{ik_x x}{\bf\hat{x}} +  \frac{k_y^2  + Q_\parallel q_\parallel - (\frac{\omega}{c})^2}{k_y^2 + q_\parallel^2 - (\frac{\omega}{c})^2}k_ye^{ik_y y}e^{i \chi}{\bf\hat{y}} \\ &+ \bigg( \frac{k_x^2 q_\parallel + Q_\parallel q_\parallel^2 - Q_\parallel(\frac{\omega}{c})^2 }{k_x^2 + q_\parallel^2 - (\frac{\omega}{c})^2}e^{i k_x x} + \frac{k_y^2q_\parallel + Q_\parallel q_\parallel^2 - Q_\parallel (\frac{\omega}{c})^2 }{k_y^2 + q_\parallel^2 - (\frac{\omega}{c})^2}e^{i k_y y}e^{i \chi}\bigg) {\bf\hat{z}}\bigg]e^{i q_\parallel z},
    \end{aligned}
    \label{Efi_pw_superposition}
\end{equation}
\end{widetext}
where ${\bf Q}={\bf k}_i+{\bf k}_f$ and ${\bf q}={\bf k}_i-{\bf k}_f$ are the same total and recoil wave vectors defined previously.

In the case of focused beams, more specifically in the small beam-width limit, we present only the vacuum transition fields involving one unit of OAM exchange between initial and final electron scattering states. The narrow beam limit, whereby the transverse electron wave function reduces to the transverse delta function \cite{de2010optical}, is adopted in all expressions. The transition vector potential and electric field associated with the HG transition current density in Eq. (\ref{J_HG}) describing the scattering from $\psi_{1 0}(x,y,z)$ to $\psi_{0 0}(x,y,z)$ is
\begin{widetext}
\begin{equation}
    \mathbf{A}^0_{00,10}(\mathbf{x}, \omega ) = - \frac{2 i e\hbar }{m  v_icw_0}\Big[ \textrm{K}_{0}\bigg(\frac{q_{\parallel} \Delta R_0}{\gamma_i}\bigg) \hat{\mathbf{x}} + i\frac{q_{\parallel} Q_{\parallel}w_0^2 }{4\gamma_i}\frac{\Delta x}{\Delta R_0}  \textrm{K}_{1}\bigg(\frac{q_{\parallel} \Delta R_0}{\gamma_i}\bigg)\hat{\mathbf{z}} \Big] e^{i q_\parallel z},
    \label{AHG}
\end{equation}
\end{widetext}
and
\begin{widetext}
\begin{equation}
    \begin{aligned}
        \mathbf{E}^0_{00,10}(\mathbf{x}, \omega ) &= \frac{2\hbar e}{mw_0 \omega v_i} \Big[ \frac{\omega^2}{c^2}\tensor{\bf I} + \nabla \nabla \Big] \cdot \Big[\textrm{K}_0\bigg(\frac{q_{\parallel} \Delta R_0}{\gamma_i}\bigg)\hat{\mathbf{x}} + i \frac{q_{\parallel} Q_{\parallel}w_0^2 }{4\gamma_i} \frac{\Delta x}{\Delta R_0}\textrm{K}_1\bigg(\frac{q_{\parallel} \Delta R_0}{\gamma_i}\bigg)\hat{\mathbf{z}}\Big] e^{iq_\parallel z} \\
        & = \frac{\hbar e}{2mw_0 \omega v_i} \Big\{  \Big( \Big[\frac{4\omega^2}{c^2} + \frac{q_{\parallel}^2}{\gamma_i^2}(4+  q_{\parallel} Q_{\parallel}w_0^2)\frac{\Delta x^2}{\Delta R_0^2} \Big]\textrm{K}_0\bigg(\frac{q_{\parallel} \Delta R_0}{\gamma_i}\bigg) + \frac{q_\parallel}{\gamma_i}(4+  {q_{\parallel} Q_{\parallel}w_0^2 } )\frac{(\Delta x^2 - \Delta y^2)}{\Delta R_0^3} \textrm{K}_1\bigg(\frac{q_{\parallel} \Delta R_0}{\gamma_i}\bigg) \Big)\hat{\mathbf{x}} \\
        & +\frac{q_\parallel^2}{\gamma_i^2}(4+  {q_{\parallel} Q_{\parallel}w_0^2 } ) \frac{\Delta x \Delta y}{\Delta R_0^2}\textrm{K}_2\bigg(\frac{q_{\parallel} \Delta R_0}{\gamma_i}\bigg)\hat{\mathbf{y}} + \frac{i}{\gamma_i}\frac{\Delta x}{\Delta R_0}\Big[  {q_{\parallel} Q_{\parallel}w_0^2 } \frac{\omega^2}{c^2}- q_\parallel^2(1 +  {q_{\parallel} Q_{\parallel}w_0^2 }  ) \Big]\textrm{K}_{1}\bigg(\frac{q_{\parallel} \Delta R_0}{\gamma_i}\bigg) \hat{\mathbf{z}} \Big\} e^{iq_\parallel z},
    \end{aligned}
    \label{EHG}
\end{equation} 
\end{widetext} 
where $\Delta R_{0} = |\mathbf{R} - \mathbf{R}_0|$ and $\Delta x=|x-x_0|$ and $\Delta y=|y-y_0|$. By symmetry, scattering from HG $\psi_{01}(x,y,z)$ to $\psi_{0 0}(x,y,z)$ can be obtained from Eqs. (\ref{AHG}) and by (\ref{EHG}) interchanging $\hat{\bf x}$ with $\hat{\bf y},$ resulting in $\mathbf{A}^0_{00,01}({\bf x}, \omega )$ and $\mathbf{E}^0_{00,01}({\bf x}, \omega )$. If the transverse state does not change in the scattering process, as is the case in the conventional EEL signal, the vacuum vector potential and electric field associated with the current density in Eq. (\ref{J_HGneqm}) where $n'=n = 0$ and $m'=m=0$ become
\begin{equation}
        \mathbf{A}^0_{0 0 ,0 0}(\mathbf{x}, \omega ) =  \Big( \frac{ \hbar e}{m v_ic}\Big)Q_{\parallel}\textrm{K}_{0}\Big(\frac{q_\parallel \Delta R_{0}}{\gamma_i}\Big)\hat{\mathbf{z}} \,  e^{i q_\parallel z} 
\end{equation}
and
\begin{widetext}
\begin{equation}
        \mathbf{E}^0_{0 0 , 0 0}(\mathbf{x}, \omega ) 
         = \frac{\hbar e}{\omega m v_i}Q_{\parallel}\Bigg[\frac{q_\parallel^2}{\gamma_i \Delta R_{0}}\textrm{K}_{1}\Big(\frac{q_\parallel \Delta R_{0}}{\gamma}\Big)\Big(\Delta x \hat{\mathbf{x}} + \Delta y \hat{\mathbf{y}}\Big) +i\Big(\frac{\omega^2}{c^2} -q_\parallel^2\Big)\textrm{K}_{0}\Big(\frac{q_\parallel \Delta R_{0}}{\gamma}\Big) \hat{\mathbf{z}}\Bigg]e^{i q_\parallel z}.
\end{equation}
\end{widetext}
The latter is the well known classical field of a uniformly moving point electron \cite{de2008probing, de2010optical} in the nonrecoil approximation where $q_\parallel=\omega/v_i$ and $Q_\parallel=2k^i_\parallel$.

Transitions between LG states $\psi_{\ell p}(\rho, \phi ,z)$ and $\psi_{\ell' p'}(\rho,\phi ,z)$ involving one unit of OAM can be derived from the above HG transitions by linear combination. Specifically, 
\begin{equation}
\mathbf{A}^0_{\ell'=0,p'=0,\ell=\pm1,p=0}(\rho,\phi,z,\omega) = \frac{1}{\sqrt{2}}\big[\mathbf{A}^0_{00,10}({\bf x}, \omega) \pm i\mathbf{A}^0_{00,01}({\bf x},\omega )\big]
\end{equation}
and
\begin{equation}
\mathbf{E}^0_{\ell'=0,p'=0,\ell=\pm1,p=0}(\rho,\phi,z,\omega) = \frac{1}{\sqrt{2}}\big[\mathbf{E}^0_{00,10}({\bf x}, \omega ) \pm i\mathbf{E}^0_{00,01}({\bf x}, \omega )\big].
\end{equation}
All electric fields introduced in the Appendix are coded within the $e$-DDA and can be used to calculate the presented EEL observables in both wide field and focused beam limits.

\subsection{Observables Under Interchange of Initial and Final States}

The EEL rate $w_{fi}$ is proportional to $\int d\mathbf{x} \, d\mathbf{x}' \mathbf{J}^{*}_{fi}(\mathbf{x}) \cdot \tensor{\mathbf{G}}(\mathbf{x}, \mathbf{x}',\omega)  \cdot \mathbf{J}_{fi}(\mathbf{x}').$ By interchanging the 3D coordinates ${\bf x}$ and ${\bf x}'$, and invoking reciprocity, i.e., $G_{\alpha \beta}(\mathbf{x}, \mathbf{x}',\omega) =G_{\beta\alpha}(\mathbf{x}', \mathbf{x},\omega)$, 
\begin{widetext}
\begin{equation}
    \begin{aligned}
        \int d\mathbf{x} \, d\mathbf{x}' \big[ {\bf J}_{fi}(\mathbf{R})e^{iq_\parallel z}\big]^*_\alpha G_{\alpha \beta}(\mathbf{x}, \mathbf{x}',\omega) \big[ {\bf J}_{fi}(\mathbf{R}')e^{iq_\parallel z'}\big]_\beta 
        &=\int d\mathbf{x} \, d\mathbf{x}' \big[ {\bf J}_{fi}(\mathbf{R})\big]^*_\alpha G_{\alpha \beta}(\mathbf{x}, \mathbf{x}',\omega) \big[ {\bf J}_{fi}(\mathbf{R}')\big]_\beta e^{-iq_\parallel (z-z')}\\
        &= \int d\mathbf{x} \, d\mathbf{x}' \big[ {\bf J}_{if}(\mathbf{R})\big]_\alpha G_{\alpha \beta}(\mathbf{x}, \mathbf{x}',\omega) \big[ {\bf J}_{if}(\mathbf{R}')\big]^*_\beta e^{-iq_\parallel (z-z')} \\
        &= \int d\mathbf{x} \, d\mathbf{x}' \big[ {\bf J}_{if}(\mathbf{R}')\big]^*_\beta G_{\beta \alpha}(\mathbf{x}', \mathbf{x},\omega) \big[ {\bf J}_{if}(\mathbf{R})\big]_\alpha e^{-iq_\parallel(z-z')} \\
        &= \int d\mathbf{x} \, d\mathbf{x}' \big[ {\bf J}_{if}(\mathbf{R})\big]^*_\alpha G_{\alpha \beta}(\mathbf{x}, \mathbf{x}',\omega) \big[ {\bf J}_{if}(\mathbf{R}')\big]_\beta e^{-iq_\parallel(z'-z)} \\
        &=\int d\mathbf{x} \, d\mathbf{x}' \big[ {\bf J}_{if}(\mathbf{R})e^{i(-q_\parallel) z}\big]^*_\alpha G_{\alpha \beta}(\mathbf{x}, \mathbf{x}',\omega) \big[ {\bf J}_{if}(\mathbf{R}')e^{i(-q_\parallel) z'}\big]_\beta, 
    \end{aligned}
    \label{ratesymm}
\end{equation} 
\end{widetext}
where $\alpha,\beta=x,y,z$ and Einstein summation notation has been used. Eq. \eqref{ratesymm} can be equivalently expressed as $w_{fi}(q_\parallel) = w_{if}(-q_\parallel)$ provided $G_{\alpha \beta}(\mathbf{x}, \mathbf{x}',\omega) =G_{\beta\alpha}(\mathbf{x}', \mathbf{x},\omega)$. Said differently, interchanging the initial and final transverse states together with changing the sign of the recoil momentum wave vector $q_\parallel$ leaves the EEL observable invariant in a reciprocal medium. In the case of an isolated dipolar target at position $\mathbf{x}_d$ characterized by frequency-dependent polarizability $\tensor{\boldsymbol{\alpha}}(\omega)$, $\tensor{\mathbf{G}}(\mathbf{x}, \mathbf{x}',\omega) $ satisfies the reciprocity condition when the polarizability tensor is complex symmetric, i.e., when $\tensor{\boldsymbol{\alpha}}(\omega) = \tensor{\boldsymbol{\alpha}}^T(\omega)$. 

\subsection{Laser-Stimulated Coherent States of the Target}
Under the assumption that the stimulating laser field couples to a single target mode (labeled $\ell$), which is driven into the coherent state $|\alpha_{\ell} \rangle$, the rate at which the probing electron gains energy as the target transitions to the sum of final Fock states $| n '_\ell\rangle$ is 
\begin{equation}
    \begin{aligned}
        w_{fi}^{\textrm{gain}} & = \frac{2\pi}{\hbar c^2}\sum_{n'_\ell} \Big| \sum_{\nu}\int d\mathbf{x} \, \mathbf{J}_{fi}(\mathbf{x},t) \cdot \langle n'_\ell| \mathbf{A}_{\nu}^{(+)}(\mathbf{x}) a_{\nu} e^{-i\omega_\nu t}| \alpha_\ell \rangle \Big|^2\delta(E_f - E_i) \\
        & = \frac{2\pi}{\hbar c^2}\sum_{n'_\ell} \Big| \alpha_{\ell} \langle n'_\ell | \alpha_\ell \rangle \int d\mathbf{x} \,  \mathbf{A}_{\ell}^{(+)}(\mathbf{x}) \cdot \mathbf{J}_{fi}(\mathbf{x}) \Big|^2\delta(E_f - E_i).
    \end{aligned}
    \label{w_fi_coherent_step1}
\end{equation}
Coherent states can be written as $|\alpha\rangle = e^{-|\alpha|^2/2}\sum_{n=0}^{\infty}(\alpha^n/\sqrt{n!}) |n\rangle$, where $|n\rangle = [(a^\dagger)^n / \sqrt{n!} ]|0 \rangle$. When expressed in terms of the electric field $\mathbf{E}_{\ell}^{(\pm)}(\mathbf{x}) = \pm (i\omega_{\ell})/c \, \mathbf{A}_{\ell}^{(\pm)}(\mathbf{x})$ together with $ \big| \langle n'_\ell | \alpha_\ell \rangle\big|^2 = (|\alpha_\ell|^{2n'_\ell}/n'_\ell!)e^{-|\alpha_\ell|^2}$, Eq. (\ref{w_fi_coherent_step1}) can be cast as
\begin{equation}
    \begin{aligned}
        w_{fi}^{\textrm{gain}} & = 2\pi \bigg( \frac{|\alpha_\ell|}{\hbar \omega_{\ell}}\bigg)^2 \bigg( \sum_{n'_\ell} \frac{|\alpha_\ell|^{2n'_\ell}}{n'_\ell!} \bigg)e^{-|\alpha_\ell|^2} \Big| \int d\mathbf{x} \, \mathbf{E}_{\ell}^{(+)}(\mathbf{x}) \cdot \mathbf{J}_{fi}(\mathbf{x}) \Big|^2\delta([E_f - E_i]/\hbar) \\
        & = 2\pi \bigg( \frac{|\alpha_\ell|}{\hbar \omega_{\ell}}\bigg)^2 \Big| \int d\mathbf{x} \, \mathbf{E}_{\ell}^{(+)}(\mathbf{x}) \cdot \mathbf{J}_{fi}(\mathbf{x}) \Big|^2\delta([E_f - E_i]/\hbar).
    \end{aligned}
\end{equation}

\bibliography{references}

\end{document}